\documentclass[aoas, preprint]{imsart}

\RequirePackage[OT1]{fontenc}
\RequirePackage{amsthm,amsmath, amssymb}
\RequirePackage[colorlinks,citecolor=blue,urlcolor=blue]{hyperref}

\RequirePackage[round]{natbib}   
\usepackage{graphicx}
\usepackage{bm}
\usepackage{fixltx2e}
\usepackage[export]{adjustbox}
\usepackage{float}

\startlocaldefs
\numberwithin{equation}{section}
\theoremstyle{plain}

\endlocaldefs

\begin{document}

\begin{frontmatter}
\title{A multivariate spatial skew-$t$ process for joint modeling of extreme precipitation indexes}
\runtitle{modeling multivariate spatial extremes}

\begin{aug}
\author{\fnms{Arnab} \snm{Hazra}\thanksref{t1}\ead[label=e1]{ahazra@ncsu.edu}},
\author{\fnms{Brian J.} \snm{Reich}\thanksref{t2}\ead[label=e2]{bjreich@ncsu.edu}}
\and
\author{\fnms{Ana-Maria} \snm{Staicu}\thanksref{t2}\ead[label=e4]{astaicu@ncsu.edu}
\ead[label=u1,url]{http://www.foo.com}}

\runauthor{Hazra, Reich and Staicu}

\affiliation{King Abdullah University of Science and Technology\thanksmark{t1} \and North Carolina State University\thanksmark{t2}}


\address{Arnab Hazra\\
CEMSE Division \\
King Abdullah University of Science and Technology \\
Thuwal 23955-6900, Jeddah, Kingdom of Saudi Arabia \\
\printead{e1}}

\address{Brian J. Reich \\
Ana-Maria Staicu \\
Department of Statistics \\
North Carolina State University \\
Raleigh, North Carolina 27695 \\
USA\\
\printead{e2}\\
\phantom{E-mail:\ }\printead*{e4}}

\end{aug}

\begin{abstract}
To study trends in extreme precipitation across US over the years 1951-2017, we consider 10 climate indexes that represent extreme precipitation, such as annual maximum of daily precipitation, annual maximum of consecutive 5-day average precipitation, which exhibit spatial correlation as well as mutual dependence. We consider the gridded data, produced by the CLIMDEX project  (\url{http://www.climdex.org/gewocs.html}), constructed using daily precipitation data.
In this paper, we propose a multivariate spatial skew-$t$ process for joint modeling of extreme precipitation indexes and discuss its theoretical properties. 
The model framework allows Bayesian inference while maintaining a computational time that is competitive with common multivariate geostatistical approaches. 
In a numerical study, we find that the proposed model outperforms multivariate spatial Gaussian processes, multivariate spatial $t$-processes including their univariate alternatives in terms of various model selection criteria.
We apply the proposed model to estimate the average decadal change in the extreme precipitation indexes throughout the United States and find several significant local changes.






\end{abstract}


\begin{keyword}
\kwd{Climate change}
\kwd{Extremal dependence}
\kwd{Extreme precipitation indexes}
\kwd{Multivariate spatial skew-$t$ process}
\kwd{Separable covariance}
\kwd{Extremal trend analysis}
\end{keyword}

\end{frontmatter}

\section{Introduction}
Extreme precipitation is one of the most important climate factors  \citep{change2007fourth}, and the 
studies concerning the long-term changes in its frequency, intensity and duration are important for sustainable development. In order to assess climate change, the World Meteorological Organization (WMO) Commission for Climatology (CCl)/CLIVAR/JCOMM Expert Team on Climate Change Detection and Indices (ETCCDI) proposed a set of indexes that characterize climate extremes. Based on the daily observations of temperature and precipitation available at the National Climatic Data Center (NCDC)'s Global Historical Climatology Network (GHCN)-Daily dataset, \cite{donat2013global} derive a suite of gridded data products called GHCNDEX (\url{http://www.climdex.org/gewocs.html}) that covers 27 climate indexes of which 10 explain extreme precipitation. To draw inference about spatiotemporal trends, \cite{donat2013global} calculate linear trends using Sen's trend estimator \citep{sen1968estimates} separately for each grid location and use Mann-Kendall test of significance \citep{kendall1955rank}. The results show fewer significant changes in precipitation compared to the temperature indexes. However this analysis is questionable, as it completely ignores spatial and mutual dependence. For a proper analysis, it is imperative to analyze the indexes jointly and account for the dependence exhibited by the data.

The multivariate spatial modeling typically assumes the data follow a Gaussian process (GP) \citep{gelfand2010multivariate} due to the GP's attractive theoretical properties, easy implementation in high-dimensional and flexible models. However, GPs are criticized for modeling spatial extremes because of the asymptotic independence between any two spatial locations except for the trivial case of exact dependence \citep{davison2013geostatistics}. In the case of a multivariate GP (MGP), asymptotic dependence across the components are similarly zero. Hence, a geostatistical approach using a MGP is questionable for modeling multivariate spatial extremes in the presence of asymptotic dependence.

Literature on univariate spatial modeling of extremes spans Bayesian hierarchical models \citep{sang2009hierarchical, sang2010continuous}, copula-based approaches \citep{ribatet2013extreme, fuentes2013nonparametric} and max-stable processes \citep{reich2012hierarchical, davison2015statistics};  \cite{davison2012statistical} reviews different approaches. Max-stable processes (MSPs) are the only possible limits for renormalized block maximums when block sizes increase to infinity (Fisher-Tippett-Gnedenko theorem; \cite{smith1990max}) where the marginals are generalized extreme value (GEV) distributions. In spite of good theoretical properties of the MSPs in explaining univariate spatial extremes, real data applications are challenging. It is possible to calculate the joint density of the observations, i.e., the multivariate GEV distribution, only for a small number of spatial locations. Some less efficient techniques for approximating the full joint distribution are available in the literature, e.g., composite likelihoods \citep{padoan2010likelihood,huser2013composite}, hierarchical Bayesian model approaches \citep{thibaud2016bayesian} etc. \cite{fuentes2013nonparametric} propose a Dirichlet process mixture copula-based model where the spatial dependence between the extreme observations is modeled nonparametrically with the marginal distributions are GEV. Considering the computational burden of the MSPs, some sub-asymptotic models have been developed by \cite{huser2017bridging}.
Factor copula models based on GPs with random mean for replicated spatial data can model tail dependence and tail asymmetry \citep{krupskii2018factor}. Based on spatial skew-$t$ processes (STPs), \cite{morris2017space} propose a Bayesian spatiotemporal model for threshold exceedances.

In spite of the availability of many modeling approaches for univariate spatial extremes, statistical models for multivariate spatial extremes are scarce. Similar to the asymptotic behavior of the univariate renormalized block maximums, the only possible limits for multivariate renormalized block maximums are multivariate max-stable processes.
\cite{genton2015multivariate} study multivariate cases of the Gaussian \citep{smith1990max}, extremal-Gaussian \citep{schlather2002models}, extremal-$t$ \citep{nikoloulopoulos2009extreme} and Brown-Resnick \citep{brown1977extreme} processes mainly, from the theoretical perspective. In order to improve the forecasts of wind gusts in Northern Germany, \cite{oesting2017statistical} propose a joint spatial model for the observations and the forecasts, based on a bivariate Brown-Resnick process. \cite{vettori2018bayesian} and \cite{reich2018multivariate} extend the hierarchical max-stable process of \cite{reich2012hierarchical} to the multivariate setting. The computational burden is high for all of these methods. For example,  \cite{genton2015multivariate}, \cite{oesting2017statistical} and \cite{reich2018multivariate} apply their models with bivariate spatial data while \cite{vettori2018bayesian} analyze five variables but with only 9 spatial locations.

In this paper, we propose a class of multivariate skew-$t$ processes (MSTPs) by extending the univariate spatial skew-$t$ process of \cite{padoan2011multivariate} and \cite{morris2017space}. A skew-$t$ distribution is chosen due to its flexibility in modeling asymmetry and heavy-tailed data. We construct a spatial skew-$t$ process considering separable covariance structure across the space and across the indexes \citep{banerjee2002prediction} along with random mean and scale. We compare numerically the performances of MGP, multivariate symmetric $t$ process (MTP), MSTP and their univariate cases in trend estimation. Finally we apply MSTPs to draw inference about the long-term trends in the extreme precipitation indexes.


The paper is organized as follows. In Section \ref{data}, we describe the CLIMDEX/GHCNDEX data and conduct a preliminary analysis. The modeling using the proposed MSTPs is described in Section \ref{methodology}. In Section \ref{computation}, we discuss Bayesian computation. In Section \ref{application}, we apply our method for analyzing the CLIMDEX indexes. Section \ref{discussions} concludes and discusses several possible extensions of the MSTP model.


\section{CLIMDEX/GHCNDEX data and exploratory analysis}
\label{data}

The CLIMDEX/GHCNDEX data repository (\url{http://www.climdex.org/gewocs.html}) includes 10 indexes that represent extreme precipitation (Table \ref{table1}). Each annual index is calculated over the period of 67 years from 1951 to 2017 on the $2.5^\circ \times 2.5^\circ$ grid. The primary data source is the National Climatic Data Center (NCDC)'s Global Historical Climatology Network (GHCN) daily dataset, which includes over 29,000 individual stations globally. For each station having at least 40 years of data, annual summary measures are calculated separately and then the summaries are interpolated to a fine grid for each year and finally the summaries are aggregated to the $2.5^\circ \times 2.5^\circ$ grid cell level. Complete details are provided in \cite{donat2013global}. Recently, some of the indexes are separately analyzed by \cite{reich2018spatial} using a spatial Markov model.

\begin{table}
\centering
\begin{large}
\caption{Description of the CLIMDEX climate indexes. $P$ is daily precipitation (mm).
The table is reproduced from \url{http://etccdi.pacificclimate.org/list_27_indices.shtml}.}
\label{table1}
\begin{tabular}{ll}
\\
Abbreviation & Description \\
\hline
Rx5day       & Annual maximum of consecutive 5-day average $P$ \\
R99p         & Annual sum of $P$ when $P > 99$th percentile\\
Rx1day       & Annual maximum of $P$ \\
R95p         & Annual sum of $P$ when $P > 95$th percentile \\
R95pT        & Annual count of days when $P > 95$th percentile\\
SDII         & Annual total $P$ divided by the number of days with $P \geq 1$ mm \\
CWD          & Maximum annual number of consecutive wet days (i.e., $P \geq 1$ mm)   \\
R10mm        & Annual number of days with $P \geq 10$ mm  \\
PRCPTOT      & Annual total precipitation from days with $P \geq 1$ mm   \\
R20mm        & Annual number of days with $P \geq 20$ mm  \\
\hline
\end{tabular}
\end{large}
\end{table}

\begin{figure}[h]
\centering
\adjincludegraphics[height = 3in, width = 5in, trim = {{.0\width} {.2\width} 0 {.28\width}}, clip]{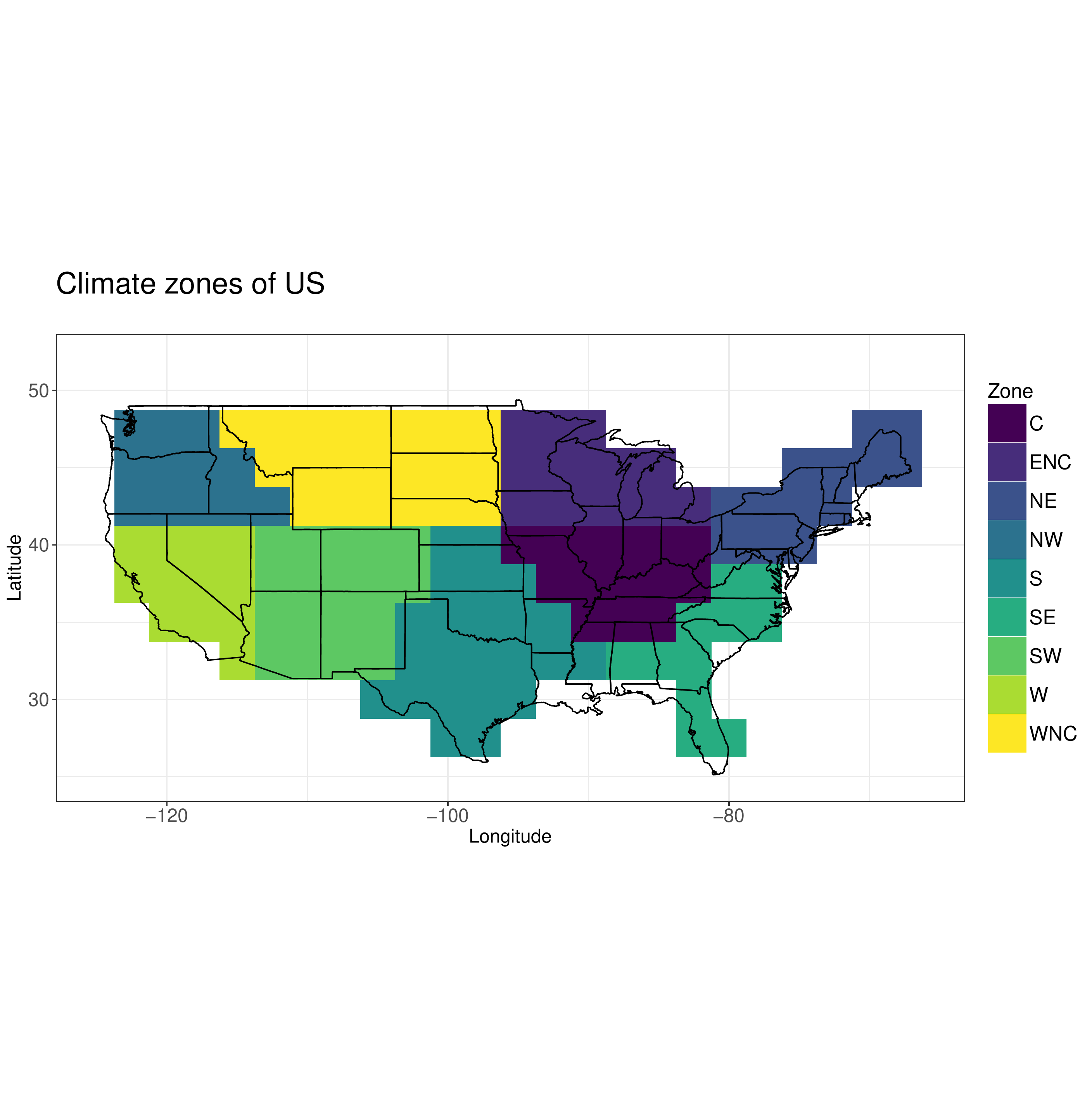}
\caption{Climate zone-wise division of the $2.5^0 \times 2.5^0$ grid covering the mainland of US according to \cite{karl1984regional}. Here N = North, S = South, E = East, W = West, C = Central.}
\label{fig1}
\end{figure}

In this paper, we consider 138 grid locations across the United States. The mainland of the United States is divided into nine climate regions according to \cite{karl1984regional} presented in Figure \ref{fig1}: Central (C), East-North-Central (ENC), North-East (NE), North-West (NW), South (S), South-East (SE), South-West (SW), West (W) and West-North-Central (WNC). We perform separate analysis for each climatically consistent region  considering the heterogenity of the climate anomalies across the regions (\url{www.ncdc.noaa.gov}).

To motivate the need for a multivariate model, we compute empirical estimates of the extremal dependence between indexes and spatial locations. The extremal dependence between two variables $Y_1$ and $Y_2$ is often measured using $\chi$-measure \citep{sibuya1960bivariate} given by
\begin{eqnarray} \label{chi}
\chi = \lim_{u \rightarrow 1} \textrm{Pr}\left[Y_1 > F_1^{-1}(u) | Y_2 > F_2^{-1}(u)\right]
\end{eqnarray}
where $F_1$ and $F_2$ are marginal distribution functions of $Y_1$ and $Y_2$ respectively. A value of $\chi$ near 1 indicates strong asymptotic dependence while $\chi = 0$ defines asymptotic independence. 
The measure $\chi$ can be estimated empirically using F-madogram \citep{cooley2006variograms} as we describe next. First, we estimate the F-madogram defined by $\nu_F = \frac{1}{2}E[ \lvert F_1(Y_1) - F_2(Y_2) \rvert ]$ based on replications of $Y_1$ and $Y_2$ and on their corresponding empirical distribution functions. Then we estimate $\chi$ by using its relationship with the F-madogram: $\chi = 2 - (1 + 2 \nu_F)/(1-2\nu_F)$. 

We can define a cross-index $\chi$-measure between two indexes $p_1$ and $p_2$ by 
\begin{eqnarray} \label{chi_cross}
\chi_{p_1, p_2} = \lim_{u \rightarrow 1} Pr\left[Y_{t p_1}(\bm{s}) > F_{Y_{t p_1}(\bm{s})}^{-1}(u) \big| Y_{t p_2}(\bm{s}) > F_{Y_{t p_2}(\bm{s})}^{-1}(u)\right]
\end{eqnarray}
where ${Y}_{tp}(\bm{s})$ denote the observation at a spatial location $\bm{s}$ and at time $t$ and $F_{Y_{tp}(\bm{s})}$ denotes the marginal distribution function of $Y_{tp}(\bm{s})$. For an index $p$, assuming the spatial process to be isotropic, the spatial extremal dependence between two locations $\bm{s}$ and $\bm{s+h}$ is
\begin{eqnarray} \label{chi_space}
\chi_{p}(h) = \lim_{u \rightarrow 1} Pr\left[Y_{tp}(\bm{s + h}) > F_{Y_{tp}(\bm{s+h})}^{-1}(u) | Y_{tp}(\bm{s}) > F_{Y_{tp}(\bm{s})}^{-1}(u)\right]
\end{eqnarray}
where $h = \Vert \bm{h} \Vert$, the Euclidean distance between $\bm{s}$ and $\bm{s+h}$.

\begin{figure}[ht]
\centering
\adjincludegraphics[height = 3in, width = 2.8in, trim = {{.09\width} {.05\width} 0 {.05\width}}, clip]{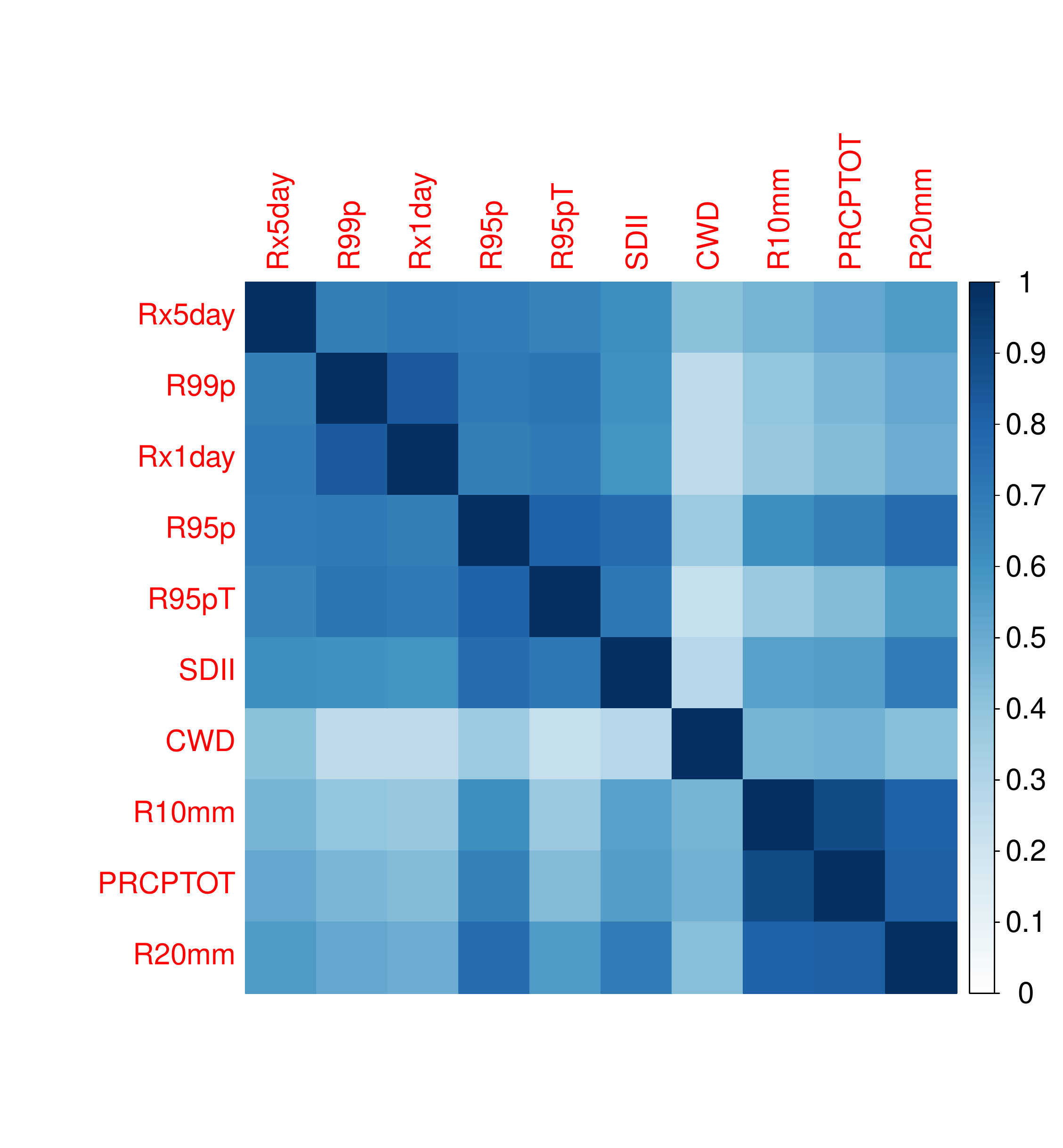}
\adjincludegraphics[height = 2.5in, width = 2.5in, trim = {{.0\width} {.0\width} 0 {.0\width}}, clip]{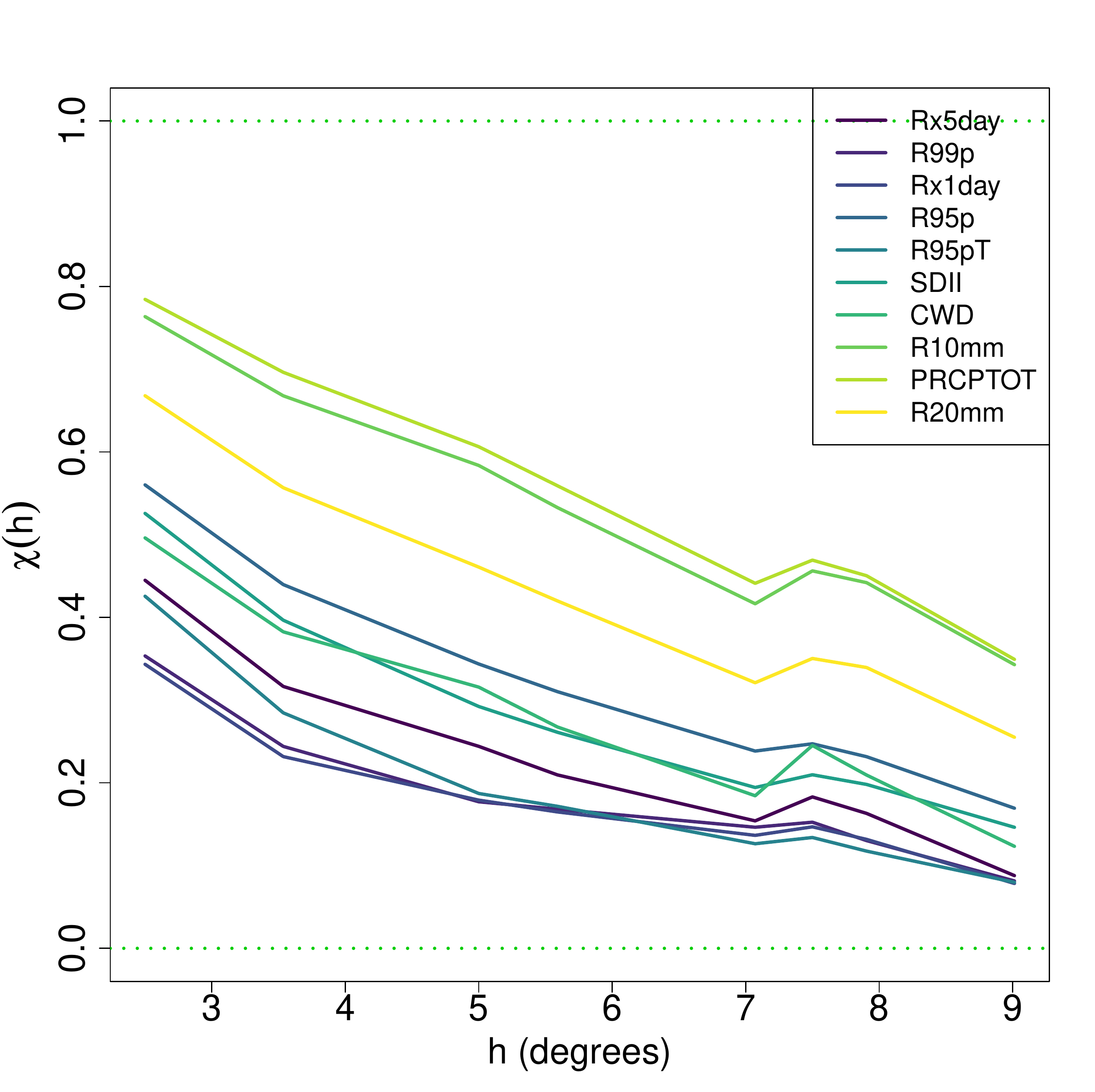}
\caption{Estimated extremal dependences between different CLIMDEX indexes (left panel). The spatial extremal dependences for each CLIMDEX index (right panel).}
\label{fig2}
\end{figure}

The extremal dependences between different CLIMDEX indexes (first separately calculated for each grid location and then averaged across them) are provided in the left panel of Figure \ref{fig2}. The plot shows evidence of strong extremal dependence among the first six indexes as well as among the last three indexes. This motivates a joint analysis of the indexes using a multivariate spatial model. 

\begin{figure}{}
\centering
\adjincludegraphics[height = 1.5in, width = 1.5in, trim = {{.0\width} {.0\width} 0 {.0\width}}, clip]{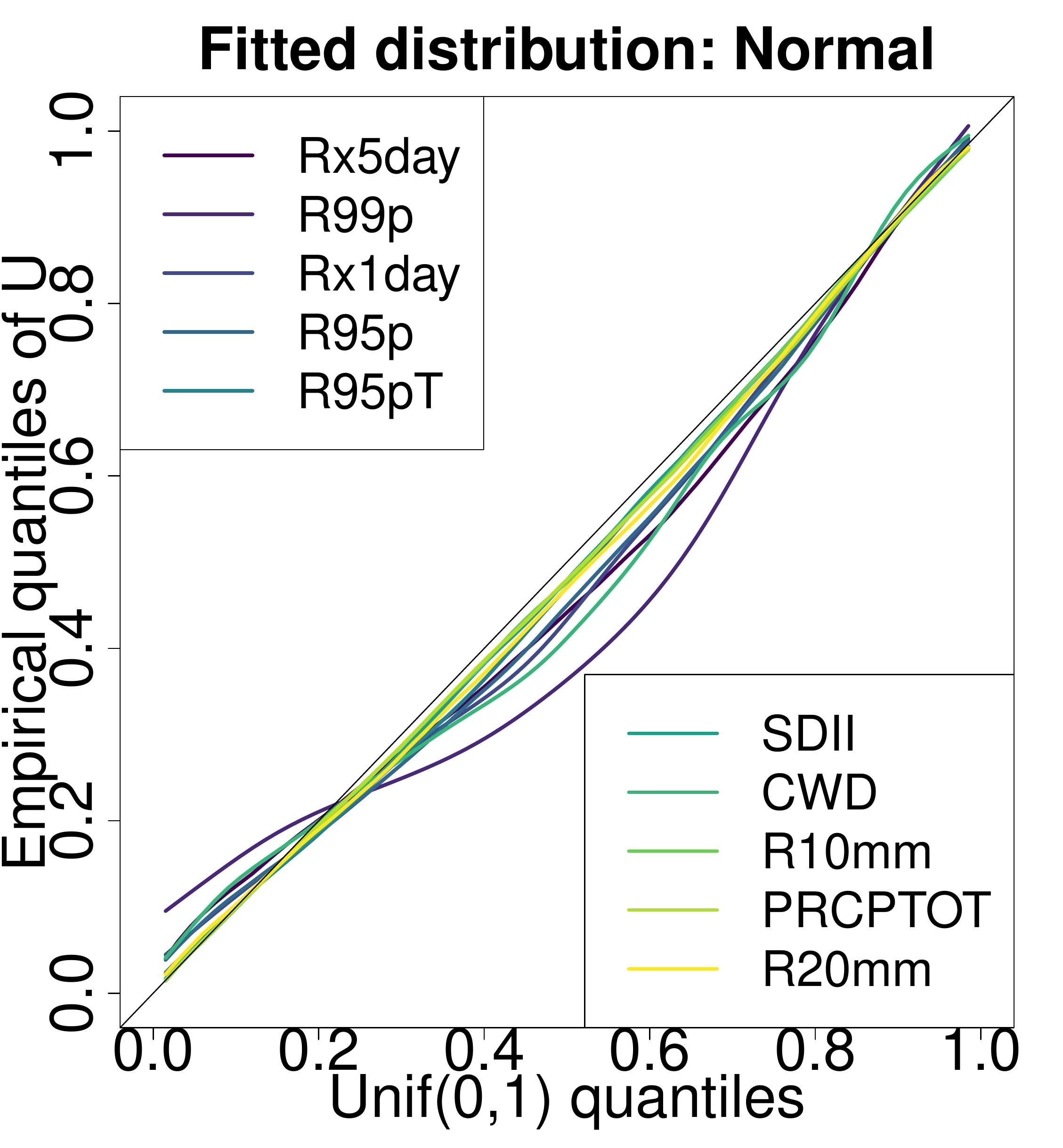}
\adjincludegraphics[height = 1.5in, width = 1.5in, trim = {{.0\width} {.0\width} 0 {.0\width}}, clip]{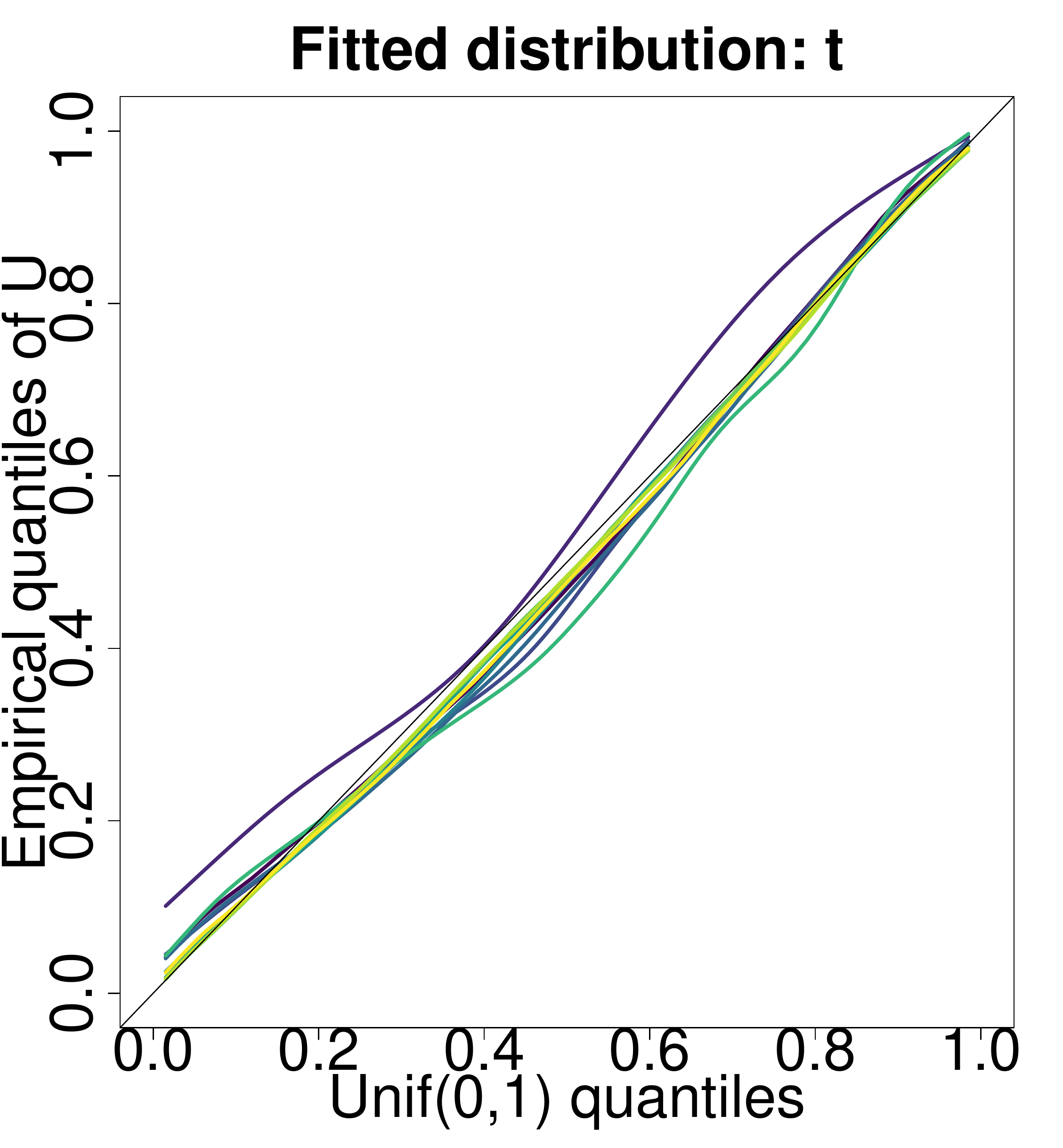}
\adjincludegraphics[height = 1.5in, width = 1.5in, trim = {{.0\width} {.0\width} 0 {.0\width}}, clip]{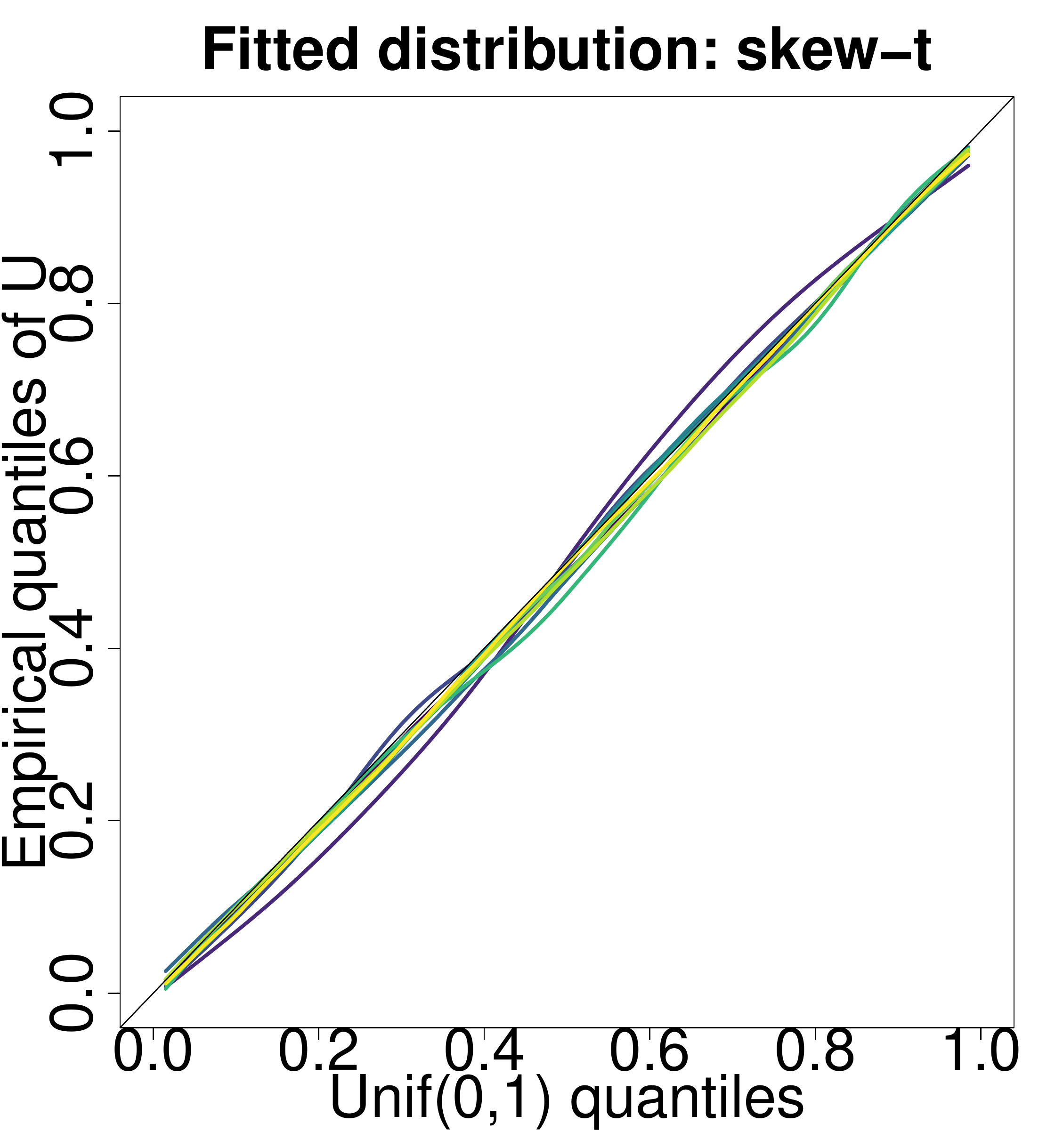}
\adjincludegraphics[height = 1.5in, width = 1.5in, trim = {{.0\width} {.0\width} 0 {.0\width}}, clip]{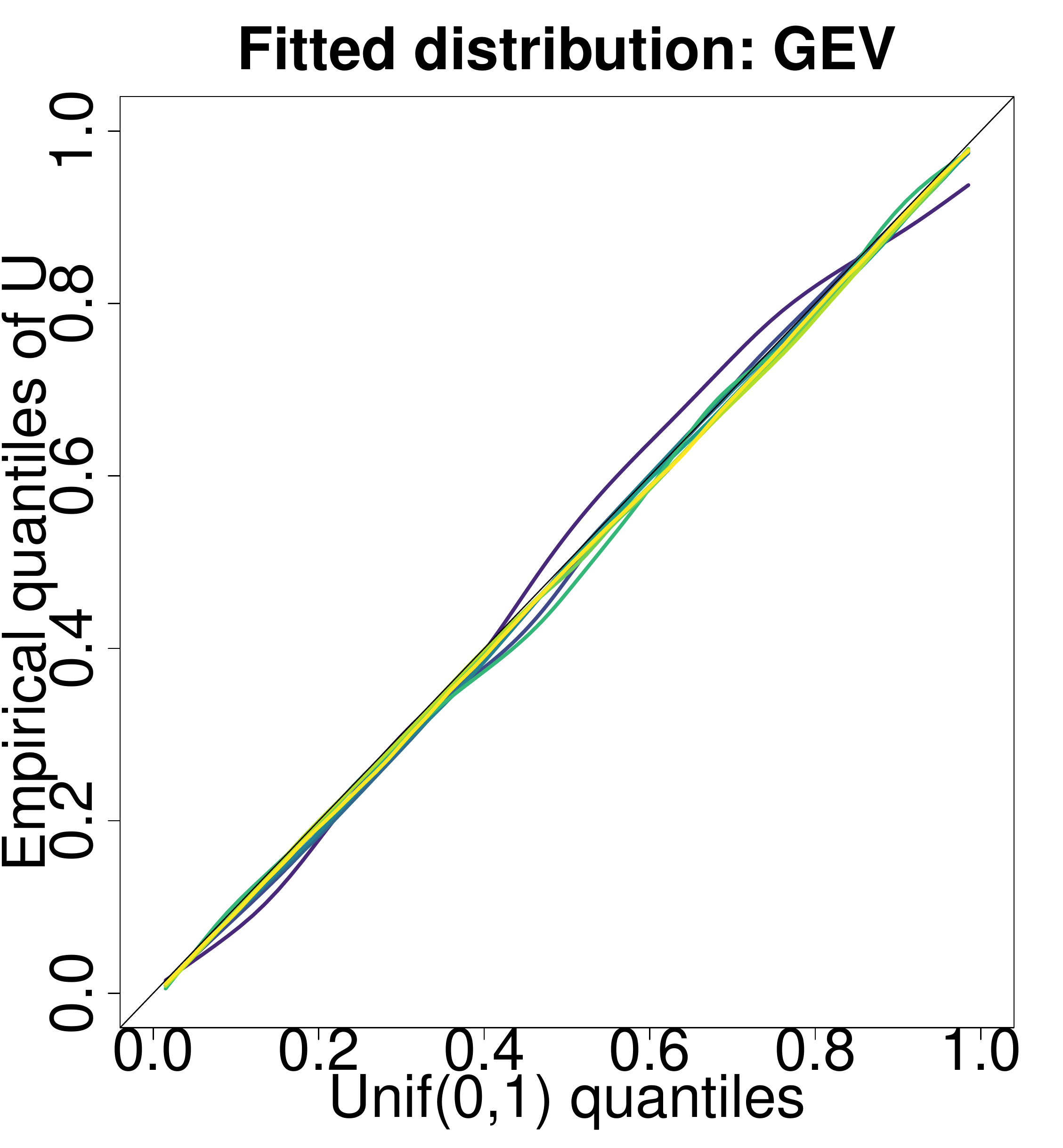}
\caption{Comparison of Unif(0, 1) quantiles versus the empirical data quantiles based on fitting different univariate distributions. The colors corresponding to different indexes are as in the first panel.}
\label{fig3}
\end{figure}

To examine the potential fit of parametric models, for each grid location and each CLIMDEX index, we separately fit normal, $t$, skew-$t$ and GEV distributions using maximum likelihood estimation. The estimated distribution functions evaluated at the observed values are expected to be uniformly distributed if the models fit the data well. In Figure \ref{fig3}, we plot the $\textrm{Unif(0,1)}$ quantiles versus the fitted data quantiles after combining across the space and time. Among the indexes, the largest deviations correspond to R99p for all four models. The normal and $t$ distributions fit worse than the skew-$t$ and GEV distributions. Comparing the performances of skew-$t$ and GEV, all the deviations above 0.9 are small for skew-$t$ while for R99p, fitting a GEV distribution leads to significant deviations. Considering the quantiles above 0.9, which are more important in return level estimation, we prefer to fit a skew-$t$ model which also has much lower computational burden.

In order to further explore the parametrization of the skew-$t$ model, we conduct a non-spatial analysis and compare the estimated parameters across indexes separately at each grid location. A random variable $Y_t$ follows a univariate skew-$t$ distribution with parameters $(\mu(t), \lambda, a, b)$ if $Y_t = \mu(t) +  \sigma_t \lvert z_t \rvert \lambda + \sigma_t \epsilon_t$ where $\epsilon_t \sim \textrm{N}(0, b)$, $z_t \sim \textrm{N}(0, 1)$ and $\sigma_t^2 \sim \textrm{Inverse-Gamma}(a/2, a / 2)$.
Suppose the observations are independently distributed as $Y_t \sim \textrm{Skew-}t\left(\mu(t), \lambda, a, b\right)$ where $\mu(t)$ denotes the trend component which is taken to be a linear combination of seven cubic B-splines (approximately one per decade; discussed in more details in Section \ref{methodology}). 
Table \ref{stpchoices} presents the mean and coefficient of variation (CV) across sites of the estimates of $\lambda$, $a$ and $b$. 
The degrees of freedom parameter $a$ have similar estimates for all the variables (varies between 10.63 and 15.56) and also small CV (varies between 0.02 and 0.26) and hence it is reasonable to assume the parameter $a$ to be constant across the variables. The estimates of $\lambda^* = \lambda / \sqrt{b}$ (we present $\lambda^*$ instead of $\lambda$ as the extremal dependence properties are determined by $\lambda^*$) and $b$ vary more across the indexes; for $\lambda^*$, the estimates vary between 1.12 for SDII and 27.53 for PRCPTOT while for $b$, the estimates vary between 0.15 for CWD and 327.29 for PRCPTOT. Thus, separate skewness and scale parameters for each climate index are needed. The CV values are relatively higher for the skewness and scale parameters as well indicating that the spatial variation of the estimates of $\lambda^*$ and $b$ are higher than $a$ within the zones and modeling the parameters $\lambda^*$ and $b$ as spatially-varying coefficients are possible choices as well though it is computationally challenging.

\begin{table}[ht]
\caption{The estimates of the parameters $\lambda^*$, $a$ and $b$ obtained by fitting univariate skew-$t$ distribution at each location. For each climate zone, the means and coefficients of variation are obtained and the values are averaged across the zones.}
\centering
\begin{tabular}{lrrrrrrrrrr}
  \hline
 & Rx5day & R99p & Rx1day & R95p & R95pT & SDII & CWD & R10mm & PRCPTOT & R20mm \\ 
  \hline
  \multicolumn{3}{l}{Mean} \\
  \hline
$\lambda^*$ & 5.88 & 10.96 & 5.62 & 11.44 & 2.24 & 1.12 & 4.25 & 1.24 & 27.53 & 2.00 \\ 
  $a$ & 12.00 & 10.63 & 12.15 & 13.76 & 13.43 & 13.35 & 12.57 & 12.99 & 15.56 & 13.18 \\ 
  $b$ & 34.17 & 27.70 & 16.10 & 87.64 & 10.81 & 0.50 & 0.15 & 7.10 & 327.29 & 2.18 \\
   \hline
   \multicolumn{3}{l}{CV} \\
    \hline
   $\lambda^*$ & 0.61 & 0.35 & 0.69 & 0.38 & 1.42 & 1.96 & 0.86 & 1.15 & 0.22 & 0.97 \\ 
  $a$ & 0.20 & 0.26 & 0.15 & 0.07 & 0.06 & 0.08 & 0.14 & 0.09 & 0.02 & 0.09 \\ 
 $b$ & 0.72 & 0.69 & 0.94 & 0.56 & 0.67 & 0.63 & 0.78 & 0.54 & 0.38 & 0.67 \\
   \hline
\end{tabular}
\label{stpchoices}
\end{table}

To motivate the need for a spatial model, suppose $\chi_p(h)$ denotes the extremal dependence between $Y_{tp}(\bm{s}_1)$ and $Y_{tp}(\bm{s}_2)$ where $Y_{tp}(\cdot)$ denotes the spatial process for the $p$-th precipitation index at time $t$ and $h = \lVert \bm{s}_1 - \bm{s}_2 \rVert$ is the Euclidean distance between $\bm{s}_1$ and $\bm{s}_2$. Assuming spatial isotropy, $\chi_p(h)$ are estimated using F-madogram for each index separately and after averaging the estimates over the pair of sites for different values of $h$ (less than the minimum of the zone-diameters) within each zone and finally averaging across the zones, the smoothed estimates are plotted in the right panel of Figure \ref{fig2}. The $\chi_p(h)$ has an overall deceasing trend for each $p$ though none of them drop to zero for the highest value of $h$ considered. At $h = 2.5$ (the smallest distance between two grid locations), the strongest extremal dependence corresponds to PRCPTOT ($\chi = 0.78$) while the weakest extremal dependence corresponds to Rx1day ($\chi = 0.34$). Therefore, we consider models that allow for a different degree of extremal dependence for each index.

\section{Methodology}
\label{methodology}

In this section, we propose a multivariate spatial skew-$t$ process (MSTP) motivated by the exploratory analysis in Section \ref{data} and then introduce a measure of multivariate spatial extremal dependence followed by the discussion of the properties of the proposed model.  
Univariate skew-$t$ processes (STPs), developed by \cite{padoan2011multivariate}, are a class of models that allow heavy-tailed and asymmetric marginal distributions and asymptotic spatial dependence. A STP is built in \cite{morris2017space} by location-scale mixing of Gaussian processes (GPs), follows from the ideas of additive processes \citep{azzalini2003distributions, azzalini2014skew}.

\subsection{Multivariate spatial skew-\texorpdfstring{$t$}{Lg} process}
A multivariate spatial skew-$t$ process is constructed in such a way that at any spatial location, the vector of observations (length is $P$, in our application, $P = 10$, the number of CLIMDEX indexes) follows a $P$-variate skew-$t$ distribution and considering $n$ spatial locations, the observations jointly follow a $nP$-variate skew-$t$ distribution.

Let $\mathbf{Y}_t(\bm{s}) = [{Y}_{t1}(\bm{s}), \ldots, {Y}_{tP}(\bm{s})]'$ denote the vector of observations at a spatial location $\bm{s}$ within the spatial domain of interest $\mathcal{D} \subset \Re^2$ and at time $t \in \lbrace 1, \ldots, T \rbrace$. We model $\mathbf{Y}_t(\bm{s})$ as
\begin{eqnarray} \label{model}
 \mathbf{Y}_t(\bm{s}) = \bm{\mu}_t(\bm{s}) + \sigma_t | z_t | \bm{\lambda} + \sigma_t \bm{\epsilon}_t(\bm{s})
\end{eqnarray}
where $\bm{\mu}_t(\bm{s}) = [{\mu}_{t1}(\bm{s}), \ldots, {\mu}_{tP}(\bm{s})]'$ is a multivariate spatio-temporal mean process, $z_t \overset{iid}{\sim} \textrm{N}(0, 1)$, $\sigma_t^2 \overset{iid}{\sim} \textrm{Inverse-Gamma}(a/2, a/ 2)$ and $\bm{\lambda}$ denotes the vector of skewness parameters.

To accommodate spatial dependence, the error processes $\bm{\epsilon}_t(\bm{s}) = [\epsilon_{t1}(\bm{s}), \ldots, \epsilon_{tP}(\bm{s})]'$ are assumed to follow $iid$ (over $t$) $P$-variate zero-mean spatial GPs with separable covariance structure \citep{banerjee2002prediction} where the $P \times P$ covariance matrix of $\bm{\epsilon}_t(\bm{s})$ is $\Sigma_I$ and the spatial correlation of the components of $\bm{\epsilon}_t(\bm{s})$ are assumed to follow an isotropic Mat\'ern correlation structure as follows
\begin{eqnarray} 
	\label{cov_structure}
	&& \textrm{cor}[\epsilon_{tp}(\bm{s}_1), \epsilon_{tp}(\bm{s}_2)] = \frac{\gamma}{\Gamma(\nu) 2^{\nu - 1}} \left( \frac{h}{\rho} \right)^{\nu} K_{\nu} \left( \frac{h}{\rho}  \right) + (1 - \gamma) I(h = 0)
	\end{eqnarray}
where the Euclidean distance between $\mathbf{s}_1$ and $\mathbf{s}_2$ is $h = \lVert \mathbf{s}_1 - \mathbf{s}_2 \rVert$, $\rho > 0$ is range, $\nu > 0$ is smoothness and $\gamma \in [0, 1]$ is the ratio of spatial to total variation. In (\ref{cov_structure}), $K_{\nu}$ is the Modified Bessel function of degree $\nu$ and $I(\mathbf{s}_1 = \mathbf{s}_2) = 1$ if $\mathbf{s}_1 = \mathbf{s}_2$ and 0 otherwise. For the observation locations $\bm{s}_1, \ldots, \bm{s}_n$, suppose the correlation matrix of $\bm{\epsilon}_{tp} = [\epsilon_{tp}(\bm{s}_1), \ldots, \epsilon_{tp}(\bm{s}_n)]'$ is $\Sigma_S$ for each $p$. Thus, the covariance matrix of $\bm{\epsilon}_t = [\bm{\epsilon}_t(\bm{s}_1)', \ldots, \bm{\epsilon}_t(\bm{s}_n)']'$ is $\Sigma_S \otimes \Sigma_I$. Let $\bm{\lambda}^*$ denotes the $P$-vector with its $p$-th element is $\lambda^*_p = \lambda_p \big/ \sqrt{\Sigma_I^{(p,p)}}$ where $\Sigma_I^{(p,p)}$ denotes the $(p,p)$-th element of $\Sigma_I$.

After marginalization over $z_t$ and $\sigma_t$, [matching the notations of \cite{azzalini2014skew}], the joint distributions of $\mathbf{Y}_t(\bm{s}_i)$ and $\mathbf{Y}_t$ are respectively
\begin{eqnarray} \label{stp}
\nonumber && \mathbf{Y}_t(\bm{s}_i) \sim \textrm{ST}_P (\bm{\mu}_t(\bm{s}_i), \Sigma_I + \bm{\lambda} \bm{\lambda}', \Sigma_I^{-1} \bm{\lambda}, a) \\
&& \mathbf{Y}_t \sim \textrm{ST}_{nP} (\bm{\mu}_t, \Sigma_S \otimes \Sigma_I + (\mathbf{1}_{n} \mathbf{1}_{n}') \otimes \bm{\lambda} \bm{\lambda}', (\Sigma_S^{-1} \mathbf{1}_{n}) \otimes (\Sigma_I^{-1} \bm{\lambda}), a)
\end{eqnarray}
where $\bm{\mu}_t = [\bm{\mu}_t(\bm{s}_1)', \ldots, \bm{\mu}_t(\bm{s}_n)']'$.

	
In order to perform a trend analysis, we assume that $\mu_{tp}(\bm{s})$'s are smooth functions of $t$ for each $\bm{s}$ and $p$ and we take the mean function to be 
\begin{eqnarray} \label{splineexp}
 \mu_{tp}(\bm{s}) = \sum_{l=1}^{L} \beta_{lp} (\bm{s}) B_l(t)
\end{eqnarray}
where $B_l(t)$ are known cubic B-spline functions of time defined over $[0, T]$ and $\beta_{lp} (\cdot)$ are spatially-varying spline coefficients. 
For convenience in exposition, we consider same basis functions for each $\bm{s}$ and $p$. 

Considering $\bm{\beta}_p(\bm{s}) = [\beta_{1p} (\bm{s}), \ldots, \beta_{Lp} (\bm{s})]'$ and $\bm{\beta}(\bm{s}) = [\bm{\beta}_{1} (\bm{s})', \ldots, \beta_{P} (\bm{s})']'$, we put a $LP$-variate spatial Gaussian process prior on $\bm{\beta}(\cdot)$. Similar to the spatial error process, considering the computational burden, we assume the covariance structure to be separable across the splines, indexes and space. Additionally, we assume that the spatial correlation structure and the correlation across the indexes for the components of $\bm{\beta}(\cdot)$ are same as those for the error process. The distribution of $\bm{\beta} = [\bm{\beta}(\bm{s}_1)', \ldots, \bm{\beta}(\bm{s}_n)']'$ is
\begin{eqnarray}
\bm{\beta} \sim \textrm{N}_{nLP} \left( \bm{1}_n \otimes \bm{\mu}_{\beta} \otimes \bm{1}_L , \Sigma_S \otimes \Sigma_I \otimes \Sigma_B \right)
\end{eqnarray}
where all the components of $\bm{\beta}$ for index $p$ are assumed to have mean $\mu_{\bm{\beta} p}$ with $\bm{\mu}_{\bm{\beta}} = [\mu_{\bm{\beta} 1}, \ldots, \mu_{\bm{\beta} P}]'$ 
and $\Sigma_B$ is a $L \times L$ covariance matrix that controls the correlation between the spline coefficients. Suppose the vector of the B-splines at time $t$ is denoted by $\bm{x}_t = [B_1(t), \ldots, B_L(t)]'$. Then the corresponding design matrix is $\bm{X}_t = I_{nP} \otimes \bm{x}_t$ with $\bm{\mu}_t = \bm{X}_t \bm{\beta}$.

\subsection{Model properties}

For the MSTP model (\ref{model}), the means and the covariances between the elements of $\bm{Y}_t$ (assuming $a > 2$) are
		\begin{eqnarray}
		\nonumber E\left[Y_{tp}(\bm{s}) \right] &=& \mu_{tp}(\bm{s}) + \lambda_p \sqrt{\frac{a}{\pi}} C(a), \\
		\nonumber Cov\left[Y_{tp_1}(\bm{s}_1), Y_{tp_2}(\bm{s}_2) \right] &=& \frac{a}{a - 2} \left[ \lambda_{p_1} \lambda_{p_2} + \Sigma_I^{(p_1, p_2)} r(h)  \right].
		\end{eqnarray}
where $C(a) = \Gamma\left(\frac{a - 1}{2} \right) / \Gamma\left(\frac{a}{2} \right)$, $\Sigma_I^{(p_1,p_2)}$ denotes the $(p_1,p_2)$-th element of $\Sigma_I$ and $h = \Vert \bm{s}_1 - \bm{s}_2 \Vert$, the Euclidean distance between $\bm{s}_1$ and $\bm{s}_2$. When $\lambda_{p} = 0$ for each $p$, the multivariate skew-$t$ model becomes multivariate symmetric-$t$ model and the covariance terms of the elements of $\bm{Y}_t$ are separable across the indexes and space. As $h$ increases to infinity, $r(h)$ converges to zero and  $Cov\left[Y_{tp_1}(\bm{s}_1), Y_{tp_2}(\bm{s}_2) \right] \rightarrow \frac{a}{a - 2} \lambda_{p_1} \lambda_{p_2} \neq 0$ if both $\lambda_{p_1}$ and $\lambda_{p_2}$ are nonzero. This property is undesirable for a spatial process defined over a large domain, e.g., the mainland of US. Partitioning of the spatial domain as in Figure \ref{fig1} is one way to force the cross-region covariance to converge to zero with increasing $h$. Considering a spatial location $\bm{s}$, $Cov\left[Y_{tp_1}(\bm{s}), Y_{tp_2}(\bm{s}) \right] = \frac{a}{a - 2} \left[ \lambda_{p_1} \lambda_{p_2} + \Sigma_I^{(p_1, p_2)} \right]$ and hence, the covariance between the indexes are determined by both the skewness parameters and the elements of $\Sigma_I$.

For the proposed MSTP model in (\ref{model}), a closed form of $\chi_{p_1, p_2}$ exists if $\lambda_{p_1}^* = \lambda_{p_2}^* = \lambda_{p_1, p_2}^*$ (say) where $\lambda_p^* = \lambda_p / \sqrt{\Sigma_I^{(p,p)}}$ and then
\begin{eqnarray} \label{extremal_cross}
 \chi_{p_1, p_2} = 2\frac{F_T\left( \lambda_{p_1, p_2}^*  \sqrt{\frac{2 a^{''}}{1 + r_{p_1, p_2}}}; a^{''} \right) }{F_T\left( \lambda_{p_1, p_2}^*  \sqrt{a'}; a'\right)} \bar{F}_T\left( \sqrt{  \frac{a' (1 - r_{p_1, p_2})}{1 + r_{p_1, p_2} + 2 \lambda_{p_1, p_2}^{*2}}}; a' \right)
\end{eqnarray}
where $r_{p_1, p_2} = \Sigma_I^{(p_1,p_2)} \big/ \sqrt{\Sigma_I^{(p_1,p_1)} \Sigma_I^{(p_2,p_2)}}$, $\bar{F}_T(\cdot~; a) = 1 - F_T(\cdot~; a)$ is the survival function for a Student's $t$ distribution with $a$ degrees of freedom, $a' = a + 1$ and $a^{''} = a + 2$.

For an index $p$, we calculate the spatial extremal dependence as
\begin{eqnarray} \label{extremal}
 \chi_{p}(h) = 2\frac{F_T\left( \lambda_{p}^*  \sqrt{\frac{2 a^{''}}{1 + r(h)}}; a^{''} \right) }{F_T\left( \lambda_{p}^*  \sqrt{a'}; a'\right)} \bar{F}_T\left( \sqrt{  \frac{a' (1 - r(h))}{1 + r(h) + 2 \lambda_{p}^{*2}}}; a' \right)
\end{eqnarray}
where $r(\cdot)$ denotes the Mat\'ern correlation function as in (\ref{cov_structure}) and other notations are as earlier.

In case of a multivariate spatial Gaussian process (MGP), $\lambda_p^* = 0$ for each $p$ and $a=\infty$. In that case, both the $\chi$-measures are zero and hence a MGP is unable to capture extremal dependences. For a multivariate spatial $t$ process (MTP), $\lambda_p^* = 0$ for each $p$ but $a$ is finite and hence the ratio terms in the expressions of both the $\chi$-measures become one. When $h \rightarrow \infty$, $r(h) \rightarrow 0$ and $\chi_{p}(h)$ is positive for any finite $\lambda_p^*$ and $a$, i.e., even for two locations infinitely apart, the spatial extremal nonzero which is a drawback of the proposed MSTP model in specific problems. Though considering a small spatial domain, for example, the climate zones in our data set-up, the assumption is reasonable follows from the right panel of Figure \ref{fig2}.

\section{Computation}
\label{computation}
We draw inference about the model parameters based on Markov chain Monte Carlo (MCMC) sampling. As the computation is highly dependent on the choice of the priors for the model parameters, we specify the priors first. We select conjugate priors when possible. The full posterior distributions of the model parameters are provided in the Appendix and an outline of the MCMC steps is discussed in this section. 

The full posterior of the spatially-varying spline coefficients is multivariate normal and hence they are updated using Gibbs sampling. Due to the choice of the separable covariance structure of $\bm{\beta}$, the posterior covariance is also separable and for updating the $nLP$ dimensional vector, the highest dimension of a matrix that needs inversion is $\max\lbrace n, L, P\rbrace$ which leads to efficient computation. For the vector $\bm\mu_{\bm\beta}$, we consider non-informative prior $\bm\mu_{\bm\beta} \sim \textrm{N}_P(\bm{0}, 100^2 I_P)$. For the skewness parameters, we assume $\bm\lambda \sim \textrm{N}(\bm{0}, 10^2 I_P)$. For the covariance matrices $\Sigma_I$ and $\Sigma_B$, we consider non-informative inverse-Wishart conjugate priors $\Sigma_I \sim \textrm{IW}(0.01, 0.01 I_P)$ and $\Sigma_B \sim \textrm{IW}(0.01, 0.01 I_L)$ respectively. These parameters are updated using Gibbs sampling. Considering the hierarchical model specification of the skew-$t$ process in (\ref{model}), the latent variables $z_t$ and $\sigma_t^2$ are updated using Gibbs sampling. Updating these parameters across $t$ are independent and hence can be updated in parallel. We consider a discrete uniform prior for the hyperparameter $a$ as $a \sim \textrm{DU}(0.1, 0.2, \ldots, 20.0)$ and the update step is using a straightforward sampling with masses at discrete values proportional to the joint likelihood of the latent variables $\sigma_t^2$. For the Mat\'ern correlation parameters $\rho$, $\nu$ and $\gamma$, there are no known conjugate priors and so we consider the priors $\rho \sim U(0, \lVert \mathcal{D} \rVert)$, $log(\nu) \sim \textrm{N}(-1.2, 1^2)$ and $\gamma \sim U(0, 1)$. Here $\lVert \mathcal{D} \rVert$ denotes the maximum Euclidean distance (in degrees) between two points within the grid $\mathcal{D}$ across US ($\lVert \mathcal{D} \rVert = 55.90$). These parameters are updated using Metropolis-Hastings sampling.

We run each MCMC chain for 20,000 iterations, discard first 10,000 iterations as burn-in and out of the post-burn-in samples, and then thin by keeping one in each five samples. Convergence of the chains are monitored by trace plots. The computing time for the MSTP model varies across climate regions and for the largest case Zone S, it is approximately 2 hours on a desktop with Intel Core i7-4790 3.60GHz processor and 32GB RAM. The computations for different zones can be run in parallel.


\section{Data application}
\label{application}
Based on Deviance Information Criterion (DIC; \cite{spiegelhalter2002bayesian}), Widely Applicable Information Criterion (WAIC; \cite{watanabe2010asymptotic,watanabe2013widely,gelman2014understanding}) and the average posterior standard deviation (SD) of the parameter of interest $\Delta_{p}(\bm{s}) = [\mu_{67,p}(\bm{s}) - \mu_{1,p}(\bm{s})]/6.6$, the average decadal change, we compare the proposed multivariate skew-$t$ process (MSTP) model with multivariate symmetric-$t$ process (MTP) obtained by setting $\bm{\lambda} = \bm{0}$ in (\ref{model}), multivariate Gaussian process (MGP) obtained by setting $\bm{\lambda} = \bm{0}$ and $a = \infty$ in (\ref{model}) and their univariate analogs STP, TP and GP. We allow separate $a$ parameter for each index for the univariate models STP and TP with the mean processes are modeled as in (\ref{splineexp}). We set $L=7$ along with considering cubic B-splines with equidistant knots. This specification allows approximately one spline per decade. The DIC and WAIC values (actual values are divided by the number of spatio-temporal points) and the average posterior SD of $\Delta_{p}(\bm{s})$ are provided in Table \ref{table_dic_waic}. For all the climate regions, the proposed MSTP model have smaller values for all three metrics compared to the other models (except for a single case) and hence preferred than other models.
 
\begin{table}[ht]
\begin{large}
\caption{DIC (Deviance Information Criterion), WAIC (Widely Applicable Information Criterion) and mean posterior standard deviation (SD) of the average decadal change $\Delta_{p}(\bm{s})$ based on fitting models MSTP, MTP, MGP, STP, TP and GP. The MSTP model have smaller values consistently across the zones (presented in bold). A model with smaller values are preferred. Values are presented after dividing the actual values by the number of spatio-temporal points.}
\centering
\label{table_dic_waic}
\begin{tabular}{lccccccccc}
  \hline
  \multicolumn{5}{l}{DIC} \\
  \hline
Model  & W & NW & SW & WNC & S & ENC & SE & C & NE \\ 
  \hline
GP & 49.81 & 44.93 & 38.57 & 39.24 & 53.07 & 44.45 & 54.52 & 50.19 & 46.56 \\ 
  TP & 47.33 & 42.62 & 37.97 & 38.88 & 52.28 & 43.83 & 53.73 & 49.56 & 45.34 \\ 
  STP & 47.02 & 42.40 & 38.32 & 39.77 & 53.42 & 43.92 & 54.39 & 53.72 & 49.63 \\ 
  MGP & 36.27 & 31.76 & 24.08 & 24.78 & 38.93 & 29.69 & 38.76 & 35.25 & 32.83 \\ 
  MTP & 35.18 & 30.92 & 23.76 & 24.62 & 38.55 & 29.44 & 38.46 & 34.73 & 31.85 \\ 
  MSTP & \textbf{34.46} & \textbf{30.35} & \textbf{23.25} & \textbf{24.16} & \textbf{38.06} & \textbf{28.83} & \textbf{38.22} & \textbf{34.26} & \textbf{31.15} \\ 
   \hline
 \multicolumn{5}{l}{WAIC} \\
\hline
 Model  & W & NW & SW & WNC & S & ENC & SE & C & NE \\  
  \hline
GP & 47.41 & 41.86 & 34.92 & 36.29 & 49.61 & 40.74 & 50.05 & 46.96 & 44.23 \\ 
  TP & 46.83 & 41.56 & 35.82 & 37.22 & 50.15 & 41.73 & 51.22 & 47.79 & 44.46 \\ 
  STP & 41.26 & 36.22 & 33.58 & 35.52 & 49.76 & 39.42 & 47.32 & 56.66 & 52.72 \\ 
  MGP & 36.19 & 31.76 & 24.09 & 24.66 & 38.82 & 29.61 & 38.81 & 35.22 & 32.64 \\ 
  MTP & 35.28 & 31.08 & 23.86 & 24.60 & 38.55 & 29.43 & 38.59 & 34.80 & 31.88 \\ 
  MSTP & \textbf{34.57} & \textbf{30.55} & \textbf{23.34} & \textbf{24.14} & \textbf{38.07} & \textbf{28.88} & \textbf{38.34} & \textbf{34.31} & \textbf{31.20} \\ 
   \hline
\multicolumn{5}{l}{Mean posterior SD of $\Delta_{p}(\bm{s})$} \\
\hline   
 Model  & W & NW & SW & WNC & S & ENC & SE & C & NE \\
  \hline
GP & 3.06 & 2.61 & 1.64 & 1.74 & 4.08 & 2.33 & 3.96 & 3.75 & 3.17 \\ 
  TP & 2.56 & 2.25 & 1.64 & 1.96 & 3.91 & 2.30 & 3.75 & 3.79 & 2.91 \\ 
  STP & 1.87 & \textbf{1.68} & 1.67 & 2.04 & 4.22 & 2.37 & 3.77 & 6.45 & 4.86 \\ 
  MGP & 2.49 & 2.18 & 1.13 & 1.36 & 3.64 & 1.88 & 3.33 & 3.00 & 2.74 \\ 
  MTP & 2.16 & 2.06 & 1.08 & 1.37 & 3.74 & 1.85 & 3.20 & 2.83 & 2.57 \\ 
  MSTP & \textbf{1.76} & 1.91 & \textbf{0.71} & \textbf{1.18} & \textbf{3.21} & \textbf{1.48} & \textbf{3.08} & \textbf{2.61} & \textbf{2.30} \\ 
   \hline
\end{tabular}
\end{large}
\end{table}

While we analyze the data based on the MSTP model as well as compare the other models for $L=7$ cubic B-splines in (\ref{splineexp}), we study the effect of the choices of $L$ on model performance by comparing the DIC and WAIC values for each zone with $L=5, 7$ and 10 and the values are provided in Table \ref{table_dic_waic_l}. The variations in DIC and WAIC are small across different choices of $L$. Hence, the choice of $L$ does not affect the inference and can be safely considered to be $L=7$.

\begin{table}[ht]
\begin{large}
\caption{DIC and WAIC based on fitting the MSTP model with different choices of the number of cubic B-splines. Small variation across the choices of $L$ indicates the insensitivity with the choice of $L$.}
\centering
\label{table_dic_waic_l}
\begin{tabular}{lccccccccc}
  \hline
  DIC \\
  \hline
 & W & NW & SW & WNC & S & ENC & SE & C & NE \\ 
  \hline
$L=5$ & 34.66 & 30.45 & 23.27 & 24.08 & 38.03 & 28.84 & 38.16 & 34.23 & 31.10 \\ 
  $L=7$ & 34.46 & 30.35 & 23.25 & 24.16 & 38.06 & 28.83 & 38.22 & 34.26 & 31.15 \\ 
  $L=10$ & 34.78 & 30.31 & 23.23 & 24.22 & 38.08 & 28.90 & 38.15 & 34.29 & 31.38 \\ 
   \hline
   WAIC \\
   \hline
   & W & NW & SW & WNC & S & ENC & SE & C & NE \\ 
  \hline
$L=5$ & 34.75 & 30.60 & 23.31 & 24.08 & 38.05 & 28.86 & 38.24 & 34.27 & 31.11 \\ 
  $L=7$ & 34.57 & 30.55 & 23.34 & 24.14 & 38.07 & 28.88 & 38.34 & 34.31 & 31.20 \\ 
  $L=10$ & 34.96 & 30.57 & 23.32 & 24.21 & 38.10 & 28.99 & 38.32 & 34.39 & 31.46 \\ 
   \hline
\end{tabular}
\end{large}
\end{table}

\begin{figure}{}
\adjincludegraphics[height = 0.4\linewidth, width=0.43\linewidth, trim = {{.0\width} {.0\width} {.0\width} {.0\width}}, clip]{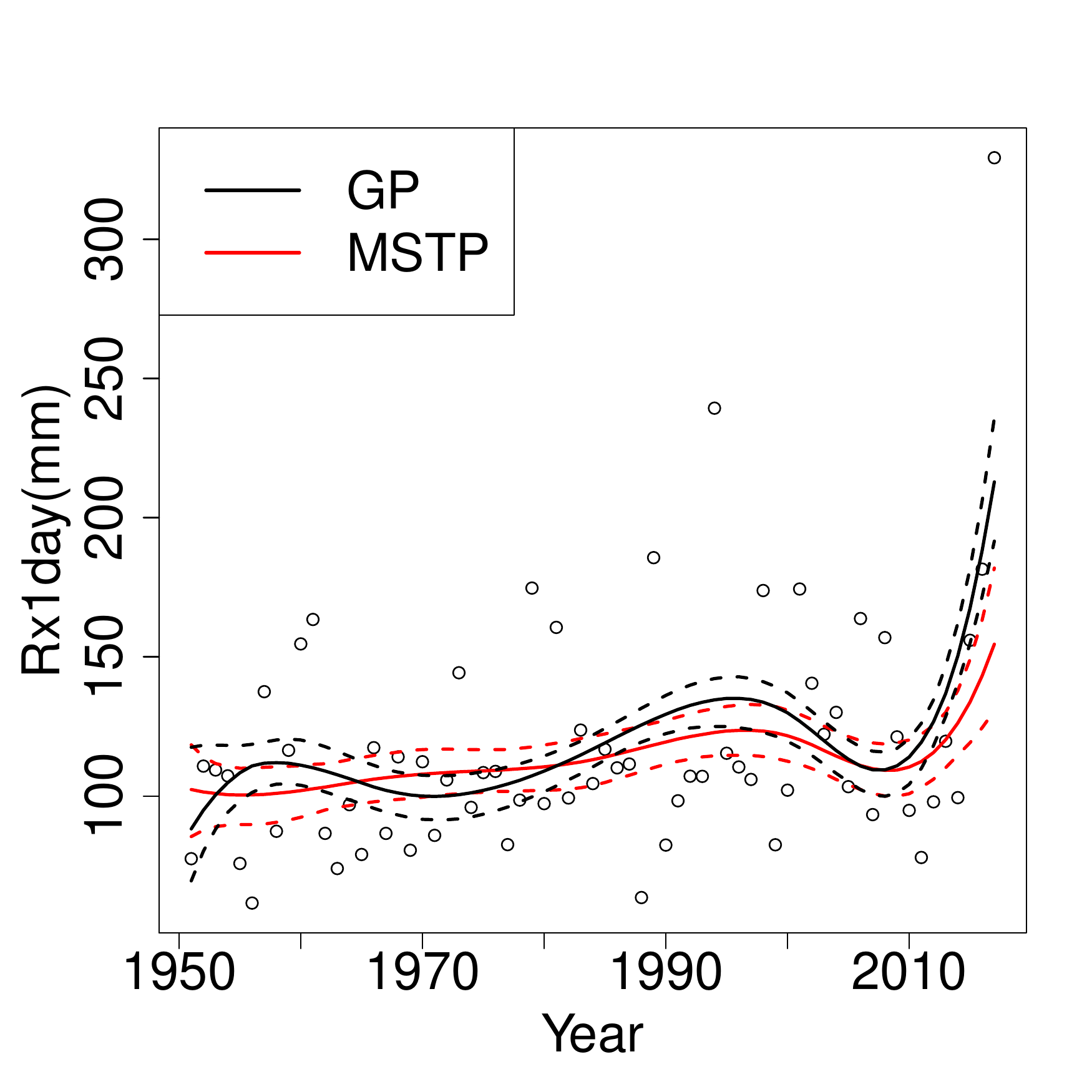}
\adjincludegraphics[height = 0.4\linewidth, width=0.45\linewidth, trim = {{.0\width} {.0\width} {.0\width} {.0\width}}, clip]{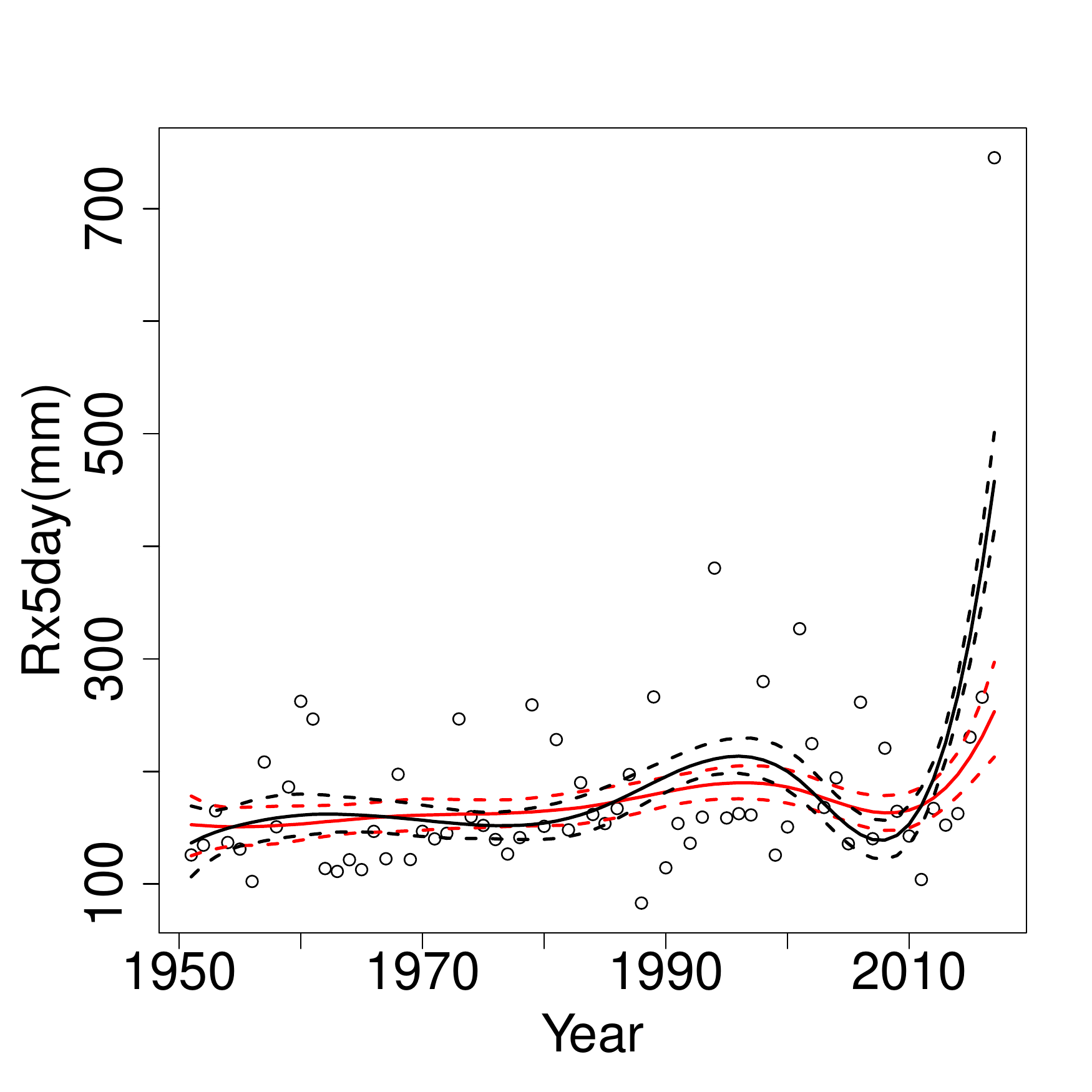}
\caption{Time series plots of Rx1day and Rx5day along with the fitted values based on the models GP and MSTP. The fitted values are based on posterior median and the 95\% posterior credible regions are presented as dashed lines.}
\label{fig_fit_houston}
\end{figure}

We illustrate the robustness of the proposed MSTP model in presence of outliers, by comparing it with GP for the specific grid point covering Houston, Texas which faced extremely high rainfall in 2017 due to hurricane Harvey. The observed values along with the estimated mean processes based on MSTP and GP are provided in Figure \ref{fig_fit_houston}. Due to the outliers for the year 2017, GP overestimates the mean for the final time point and hence the overall change between 1951-2017 as well. The average decadal change $\Delta_{p}(\bm{s})$ for Rx1day and Rx5day at that grid point are 7.90 mm and 15.22 mm respectively based on MSTP while they are 18.89 mm and 48.69 mm respectively based on GP.

\begin{figure}{}
\adjincludegraphics[height = 0.26\linewidth, width=0.45\linewidth, trim = {{.0\width} {.45\width} {.0\width} {.48\width}}, clip]{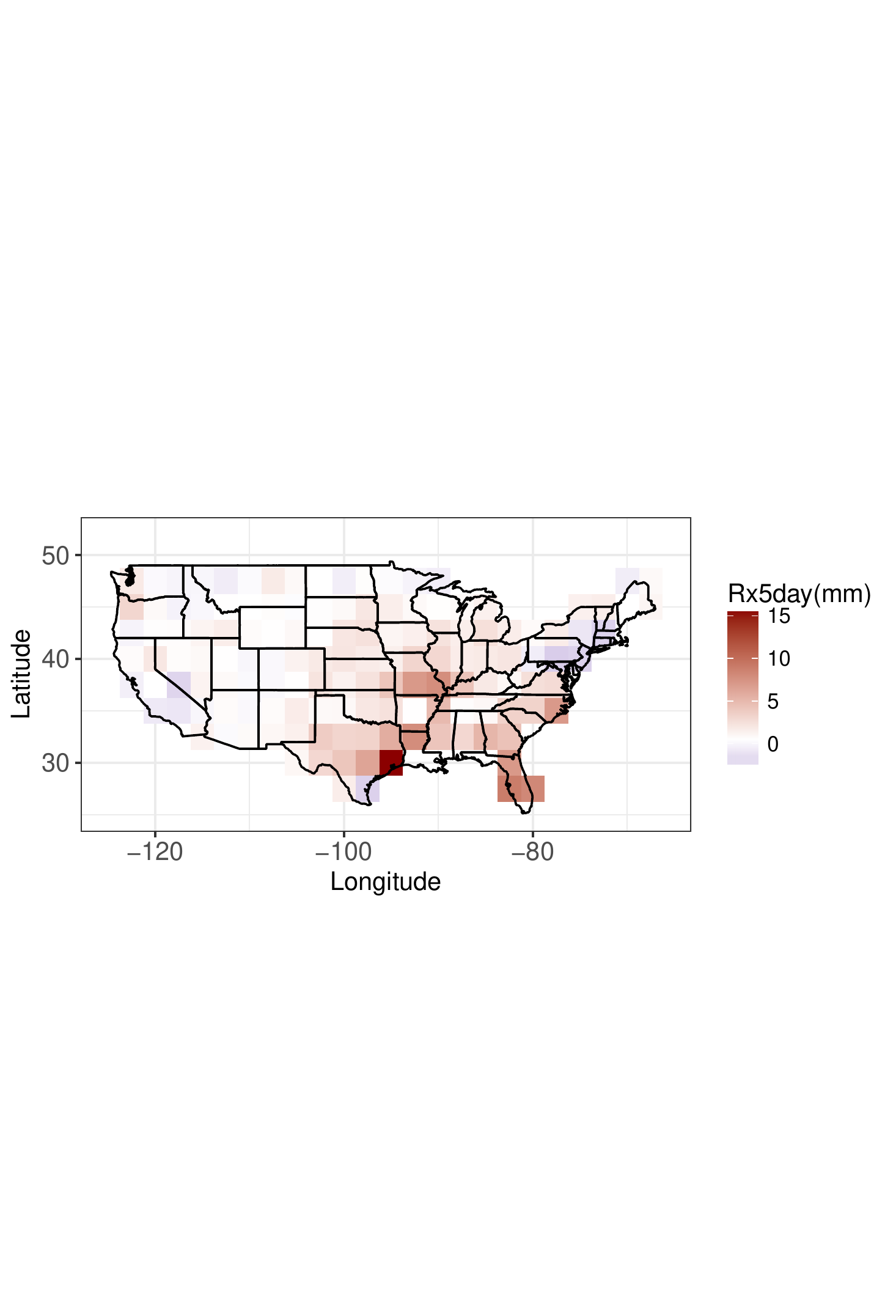}
\adjincludegraphics[height = 0.26\linewidth, width=0.45\linewidth, trim = {{.0\width} {.45\width} {.0\width} {.48\width}}, clip]{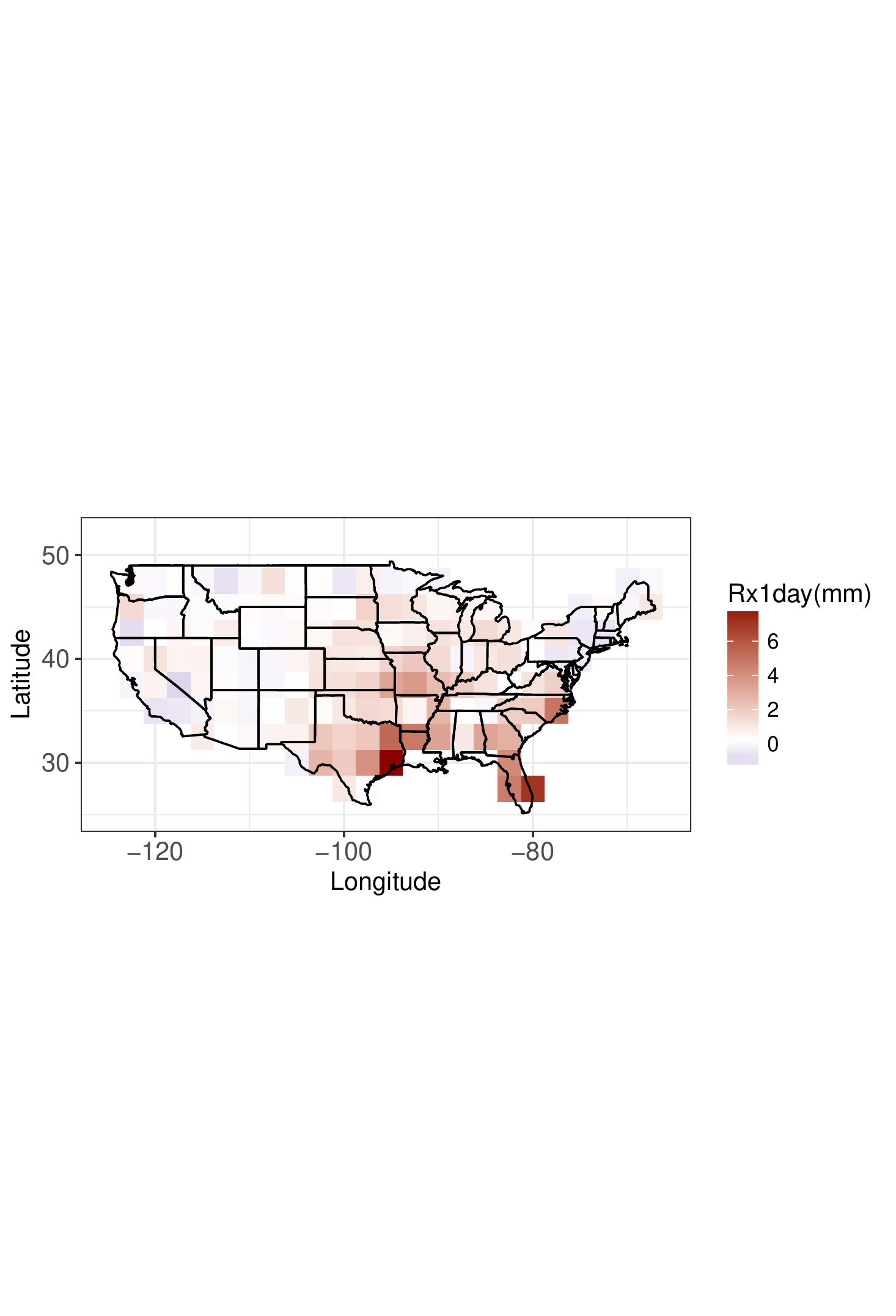} \\
\adjincludegraphics[height = 0.26\linewidth, width=0.45\linewidth, trim = {{.0\width} {.45\width} {.0\width} {.48\width}}, clip]{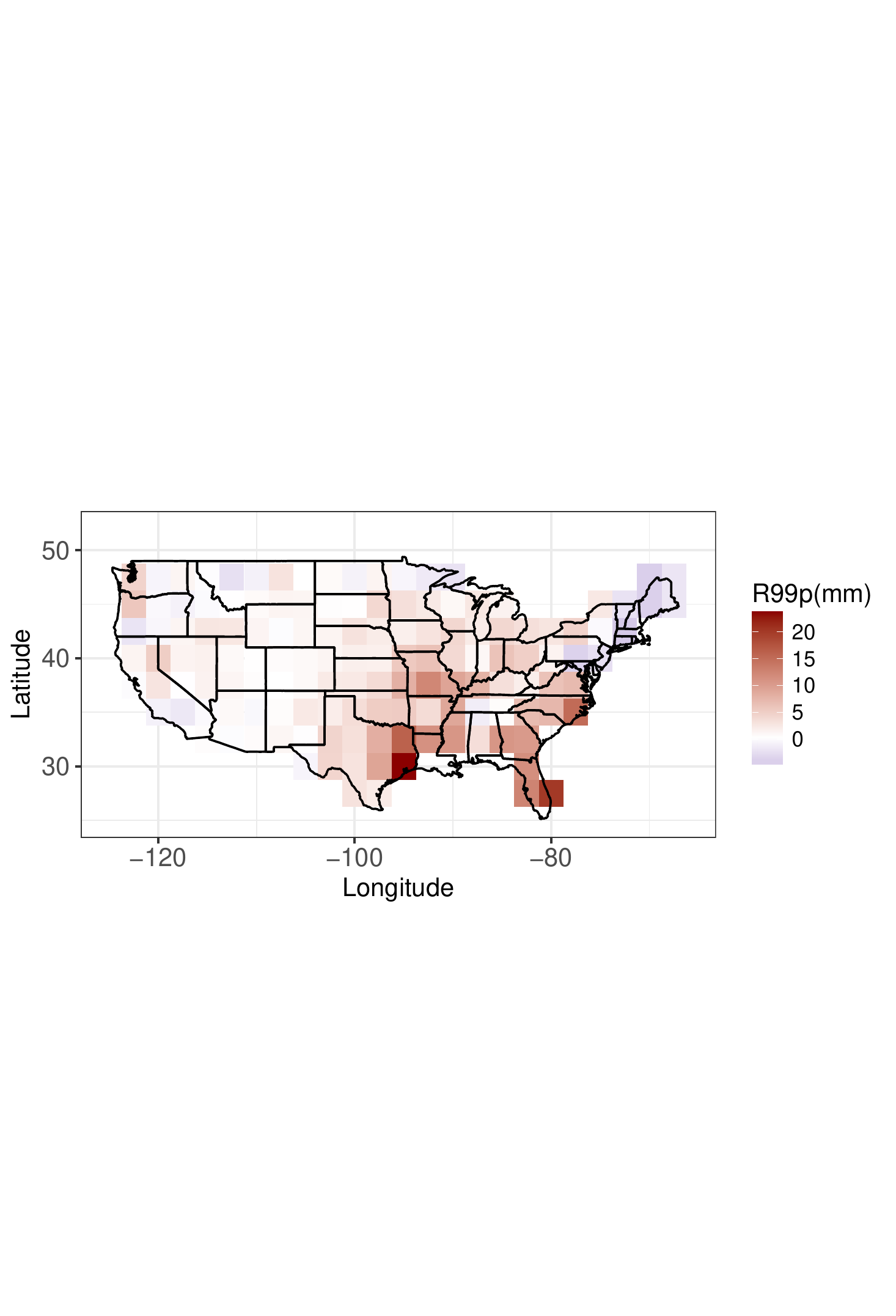}
\adjincludegraphics[height = 0.26\linewidth, width=0.45\linewidth, trim = {{.0\width} {.45\width} {.0\width} {.48\width}}, clip]{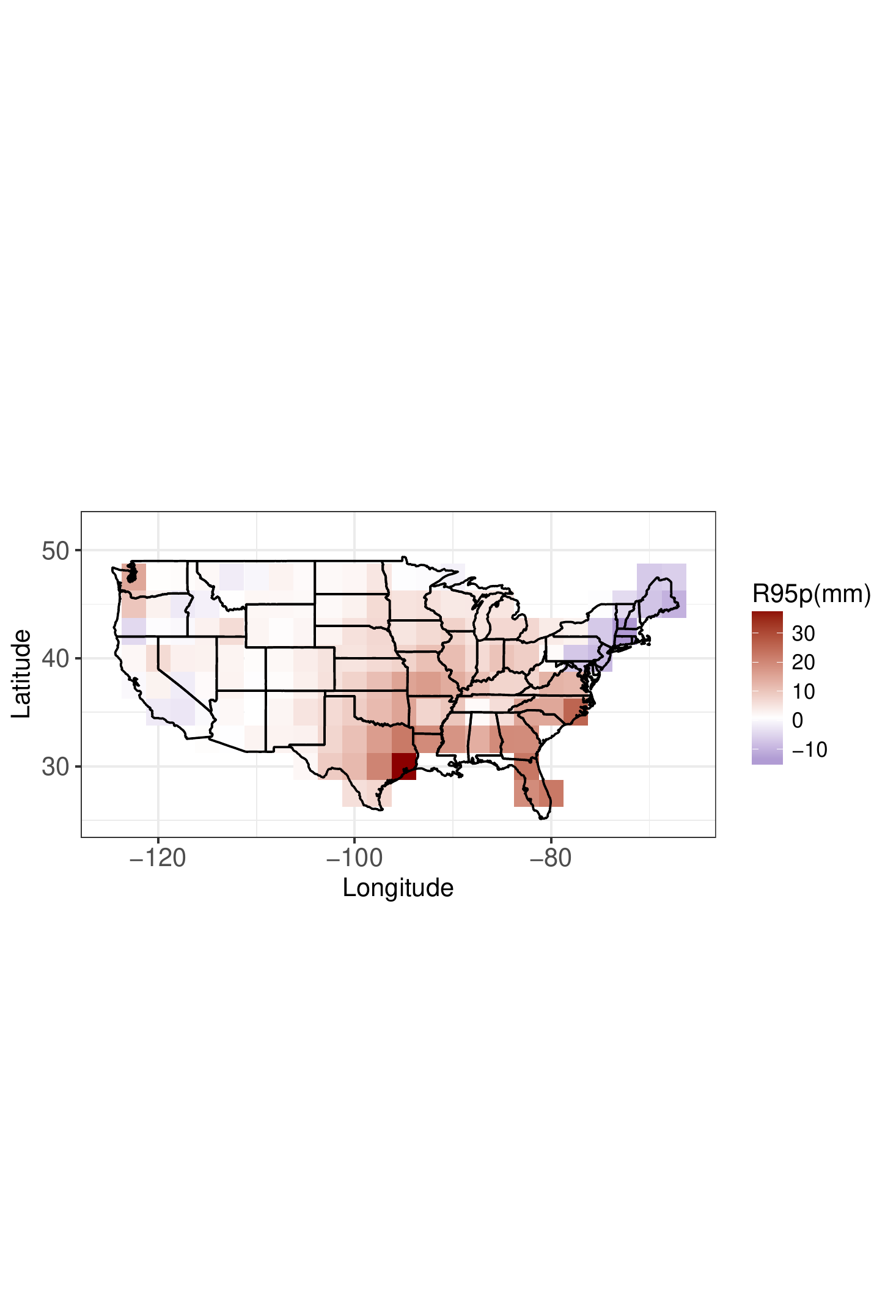} \\
\adjincludegraphics[height = 0.26\linewidth, width=0.45\linewidth, trim = {{.0\width} {.45\width} {.0\width} {.48\width}}, clip]{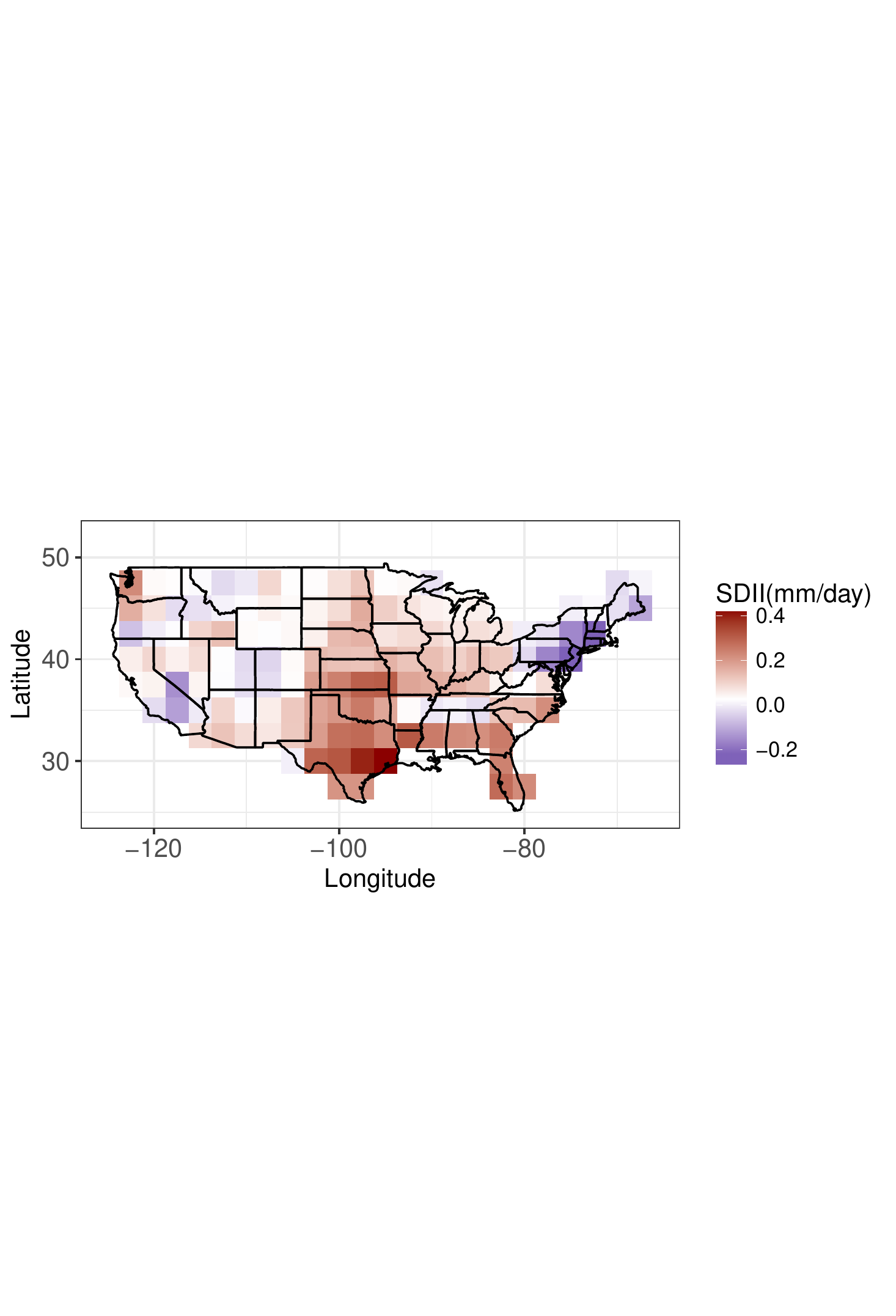}
\adjincludegraphics[height = 0.26\linewidth, width=0.45\linewidth, trim = {{.0\width} {.45\width} {.0\width} {.48\width}}, clip]{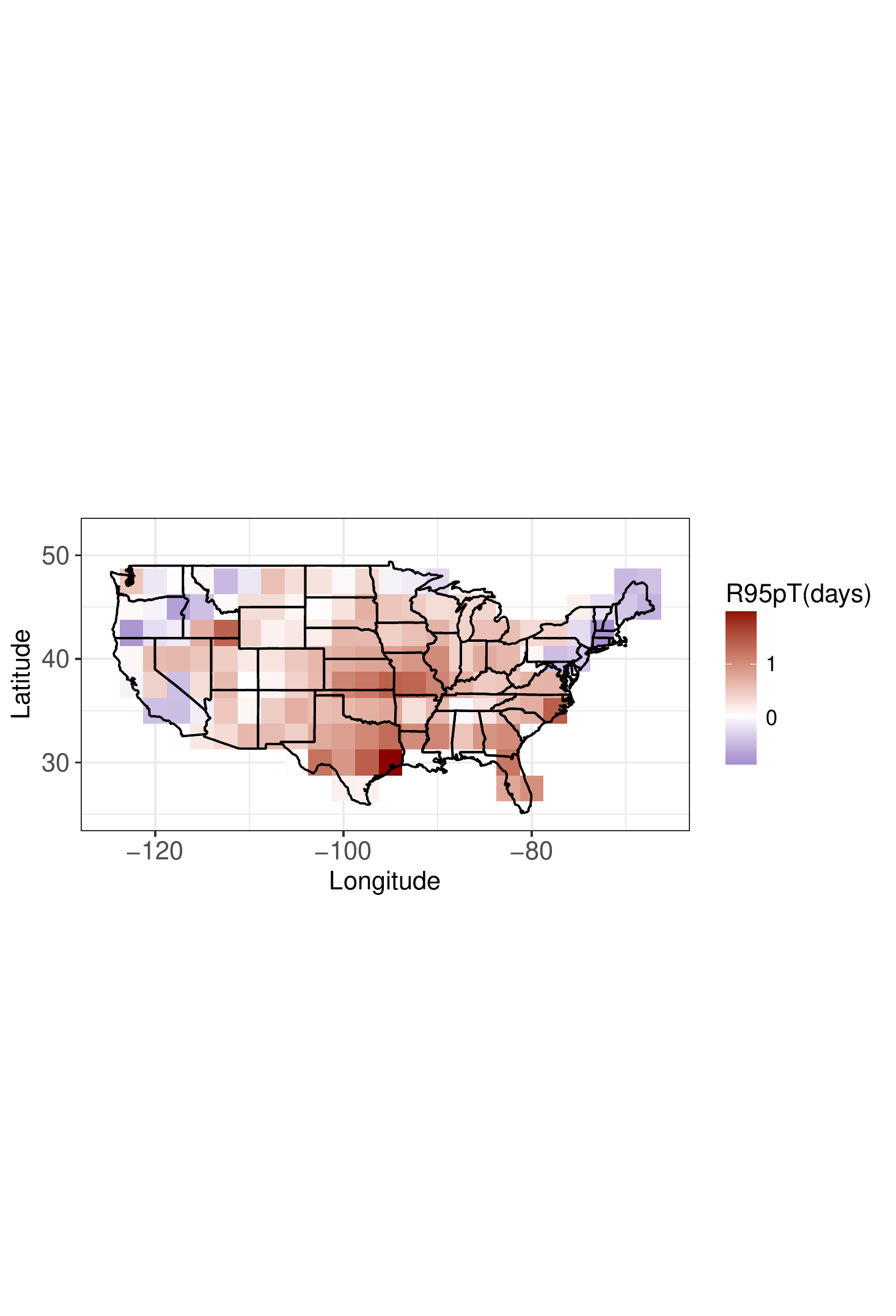} \\
\adjincludegraphics[height = 0.26\linewidth, width=0.45\linewidth, trim = {{.0\width} {.45\width} {.0\width} {.48\width}}, clip]{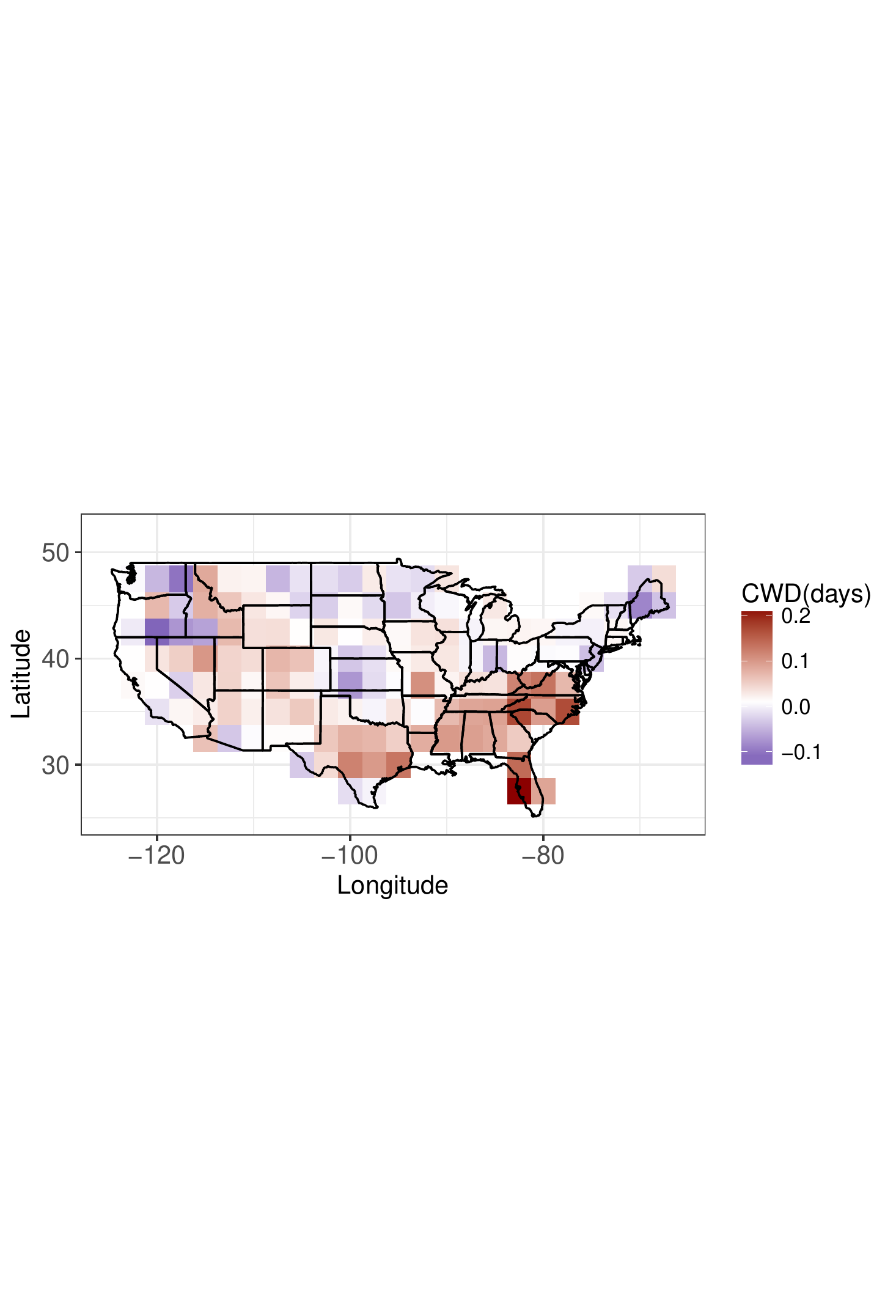}
\adjincludegraphics[height = 0.26\linewidth, width=0.45\linewidth, trim = {{.0\width} {.45\width} {.0\width} {.48\width}}, clip]{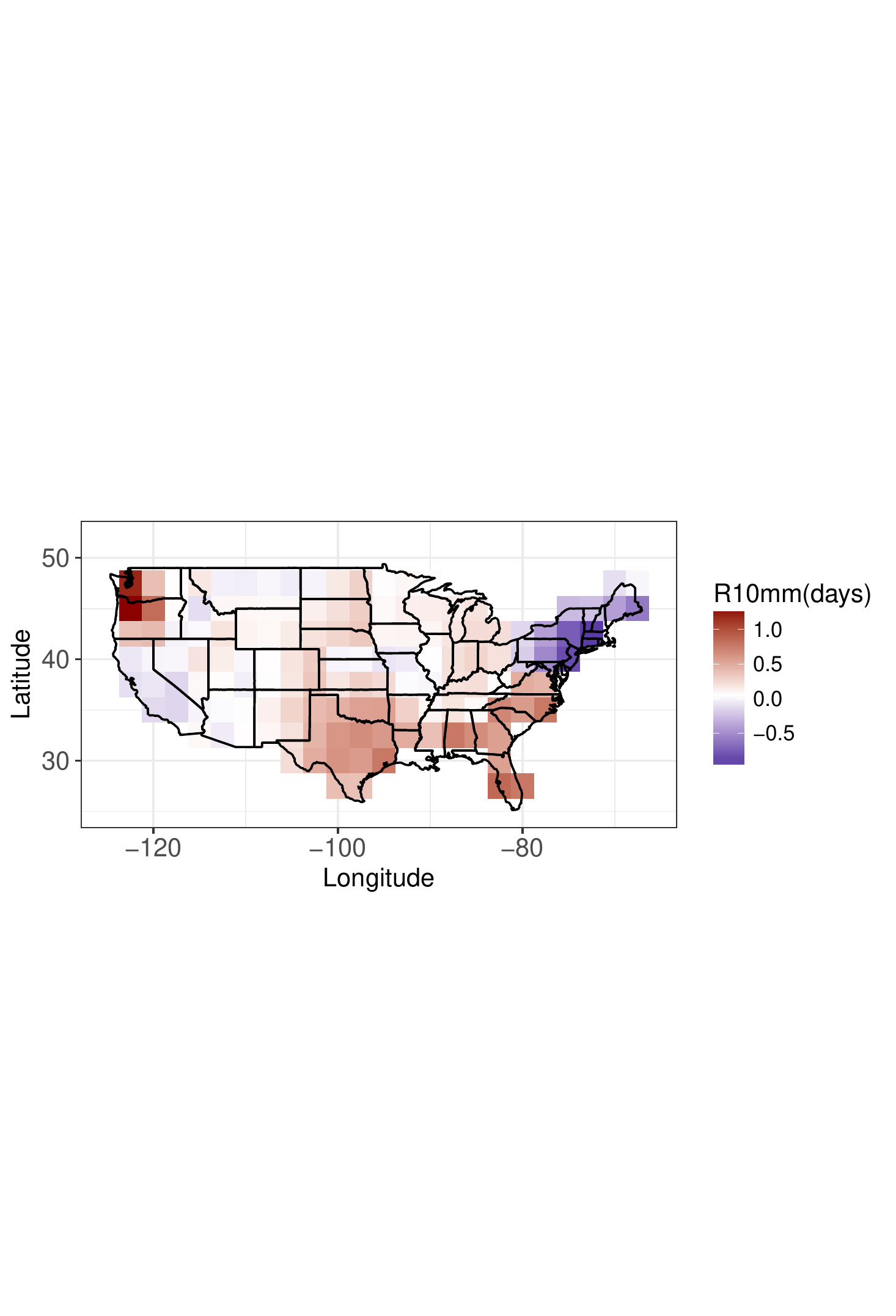} \\
\adjincludegraphics[height = 0.26\linewidth, width=0.45\linewidth, trim = {{.0\width} {.45\width} {.0\width} {.48\width}}, clip]{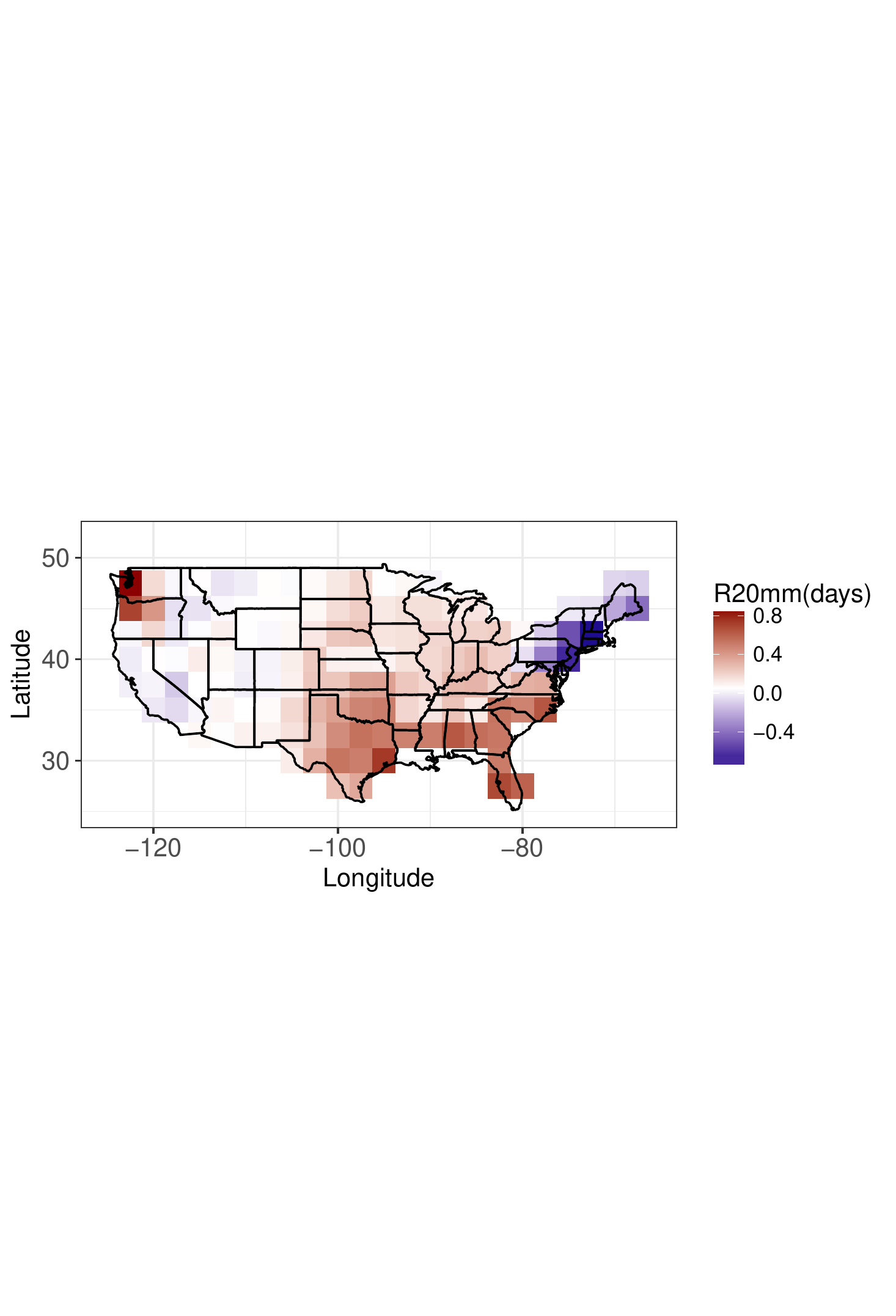}
\adjincludegraphics[height = 0.26\linewidth, width=0.45\linewidth, trim = {{.0\width} {.45\width} {.0\width} {.48\width}}, clip]{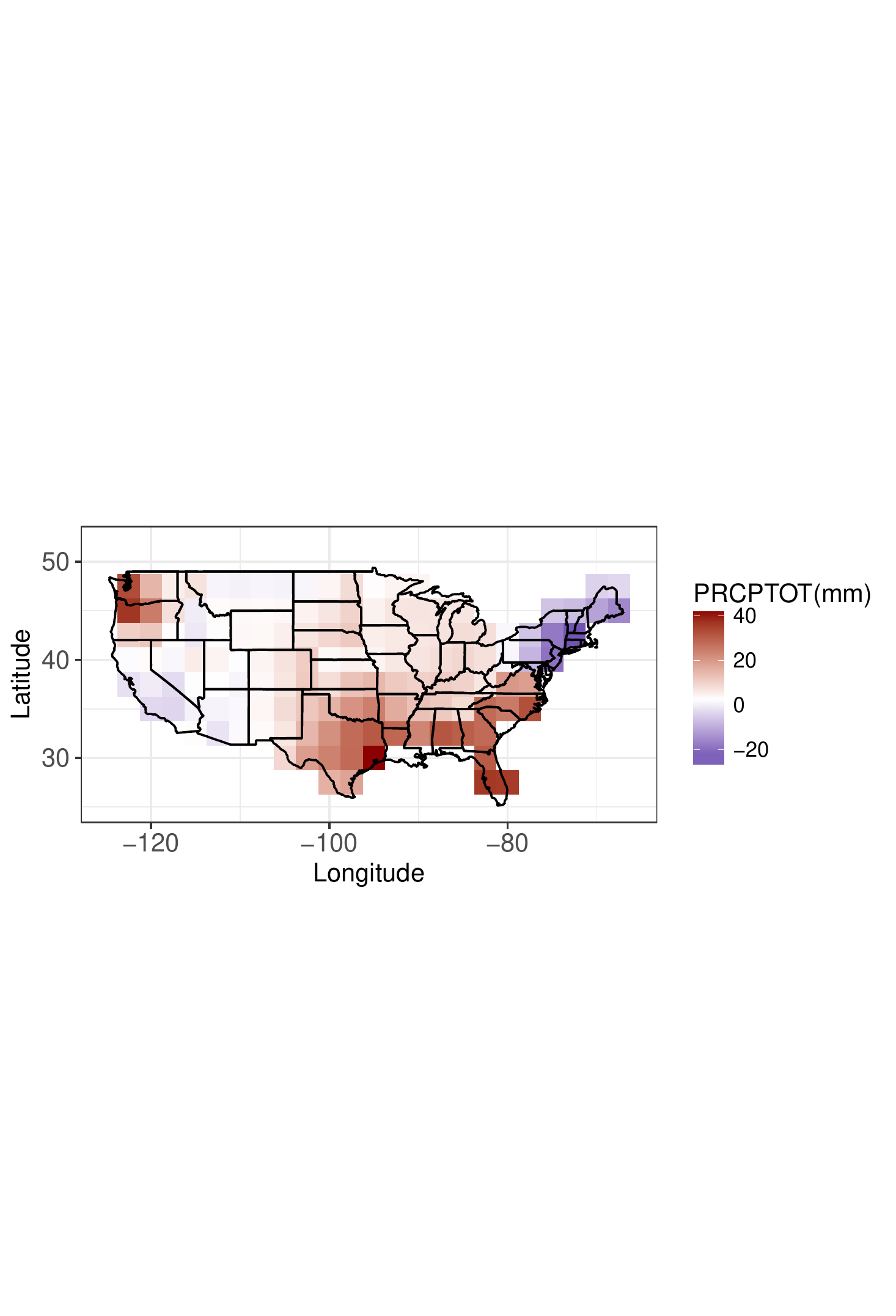}
\caption{Spatial maps of the posterior mean change per decade for the 10 precipitation indexes.}
\label{fig_betahat}
\end{figure}

Based on fitting the proposed MSTP model, we calculate the posterior mean of $\Delta_p(\bm{s})$ and the estimates the provided in Figure \ref{fig_betahat}. For all the indexes, the positive values of $\Delta_p(\bm{s})$ are observed in the South and South-East Zones. Considering consecutive 5-day precipitation, highest value corresponds to the grid point of Houston, Texas. Besides, moderate positive changes are observed in parts of Louisiana, Florida, Missouri and North Carolina. Considering annual maximum precipitation, highest positive changes are observed near the south-east of Texas and Florida. R99p has very similar spatial pattern as Rx1day. While R95p has maximum positive change of more than 30mm in the south-east Texas, negative changes are observed in the North-East zone. SDII has similar spatial pattern with R95p along with negative changes are observed in southern California. R95pT has similar spatial pattern with R95p. The positive change in CWD is maximum in the south-west of Florida followed by the eastern and western parts of North Carolina. R10mm and R20mm have very similar spatial pattern with the highest values are observed at the grid points near Seattle, Washington and the negative changes are maximum in Connecticut and New Jersey. Spatial pattern of PRCPTOT is similar to SDII along with high positive changes are observed near Seattle.

\begin{figure}{}
\adjincludegraphics[height = 0.25\linewidth, width=0.42\linewidth, trim = {{.0\width} {.15\width} {.0\width} {.15\width}}, clip]{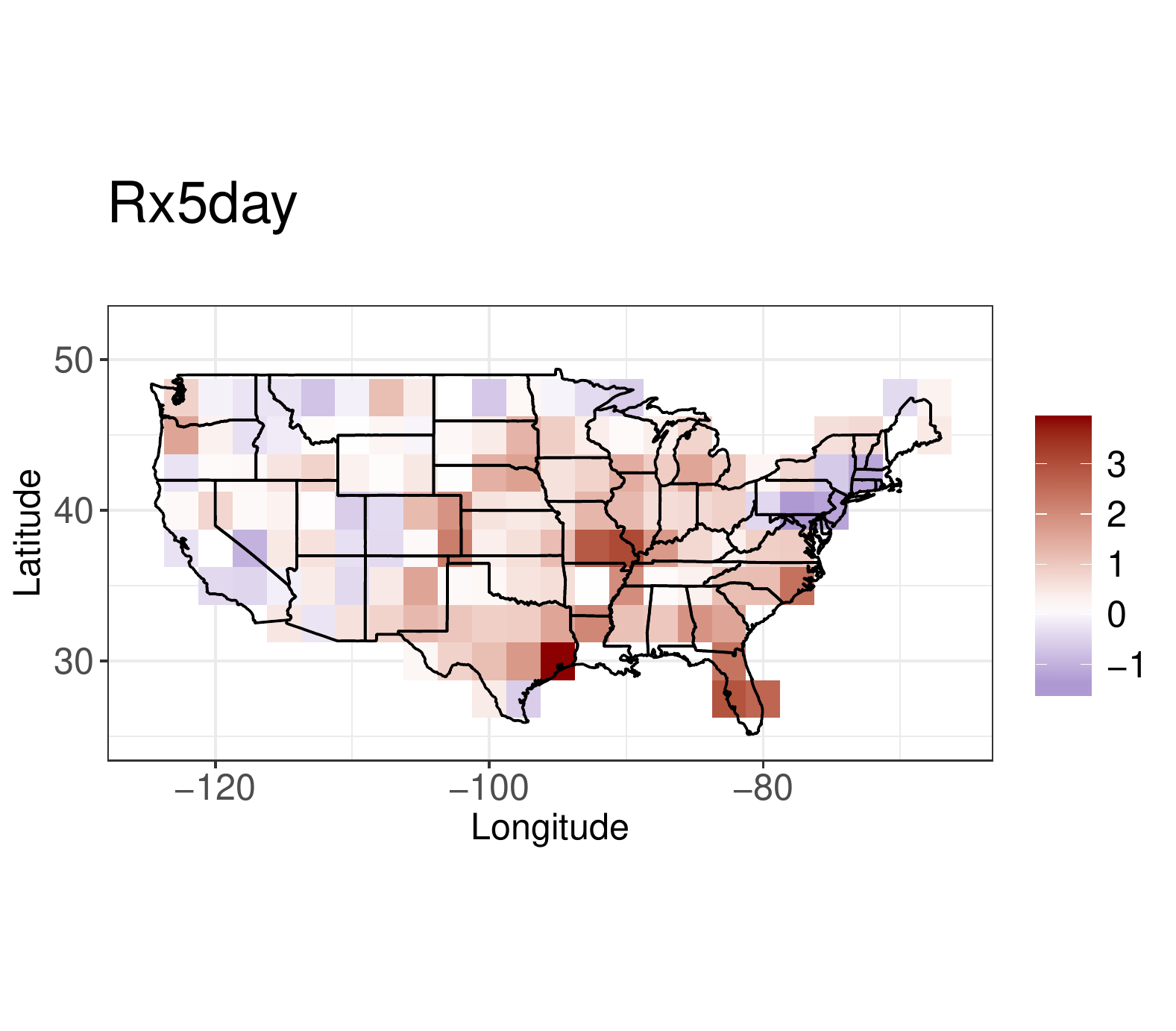} 
\adjincludegraphics[height = 0.25\linewidth, width=0.42\linewidth, trim = {{.0\width} {.15\width} {.0\width} {.15\width}}, clip]{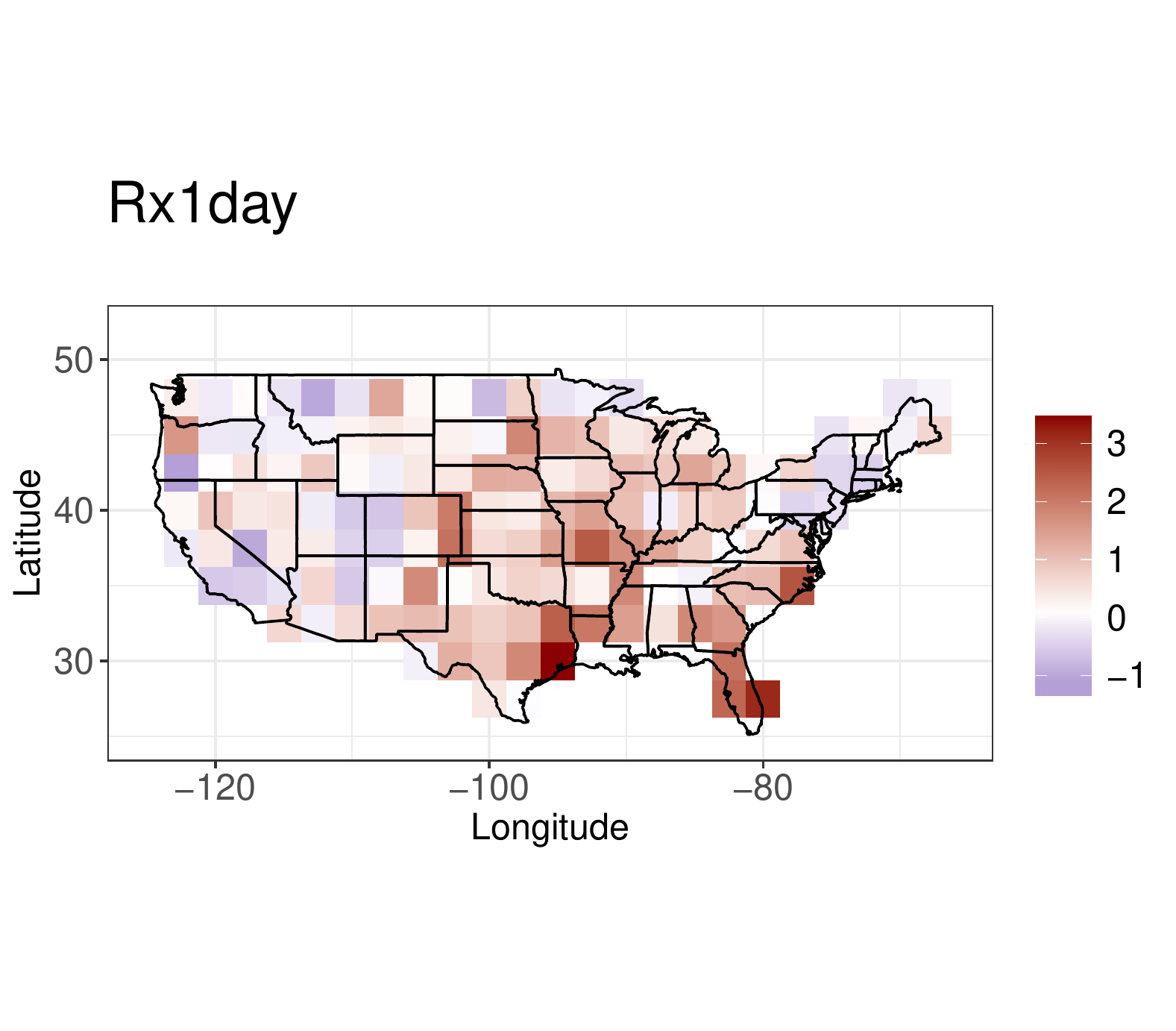} \\
\adjincludegraphics[height = 0.25\linewidth, width=0.42\linewidth, trim = {{.0\width} {.15\width} {.0\width} {.15\width}}, clip]{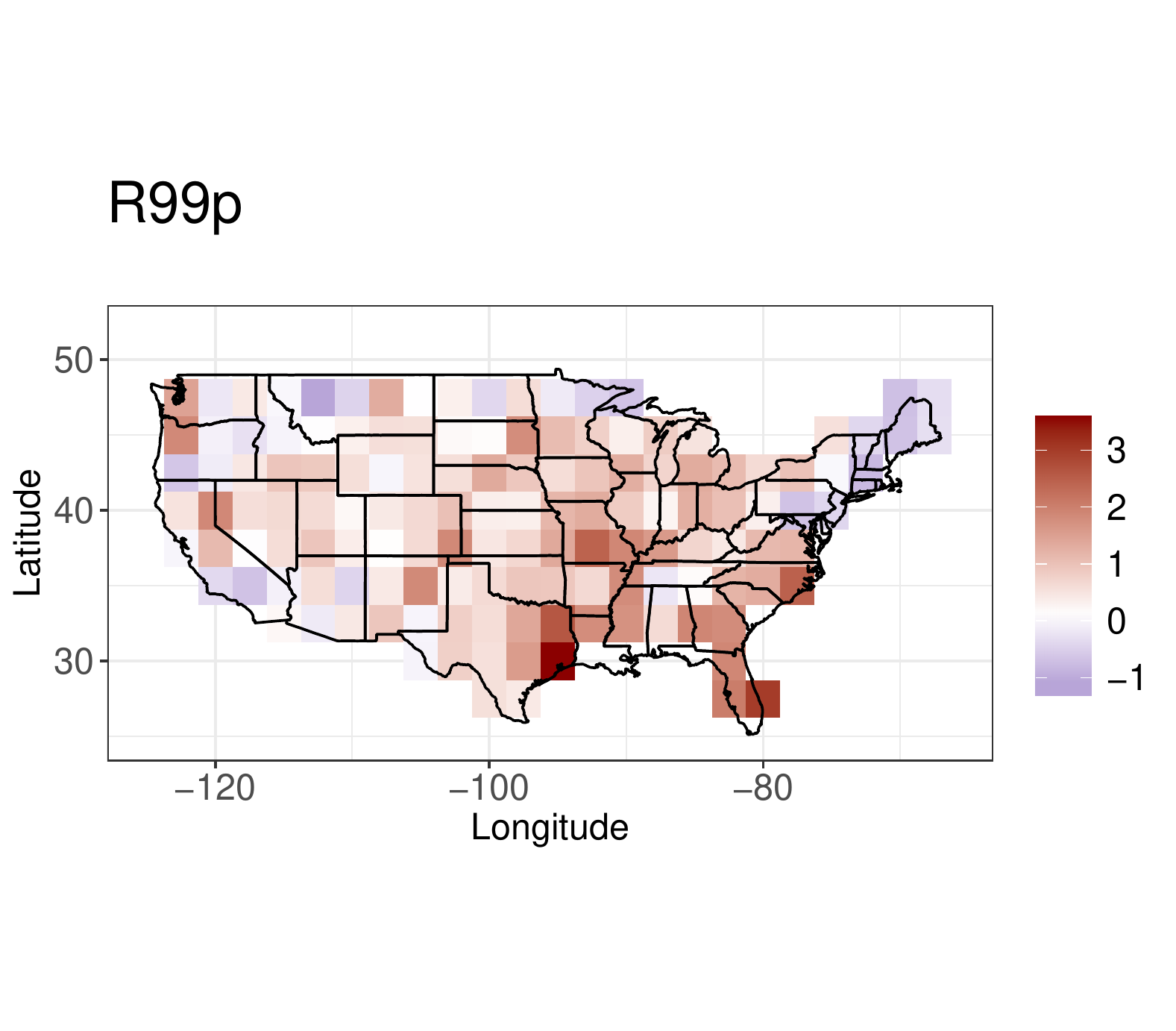} 
\adjincludegraphics[height = 0.25\linewidth, width=0.42\linewidth, trim = {{.0\width} {.15\width} {.0\width} {.15\width}}, clip]{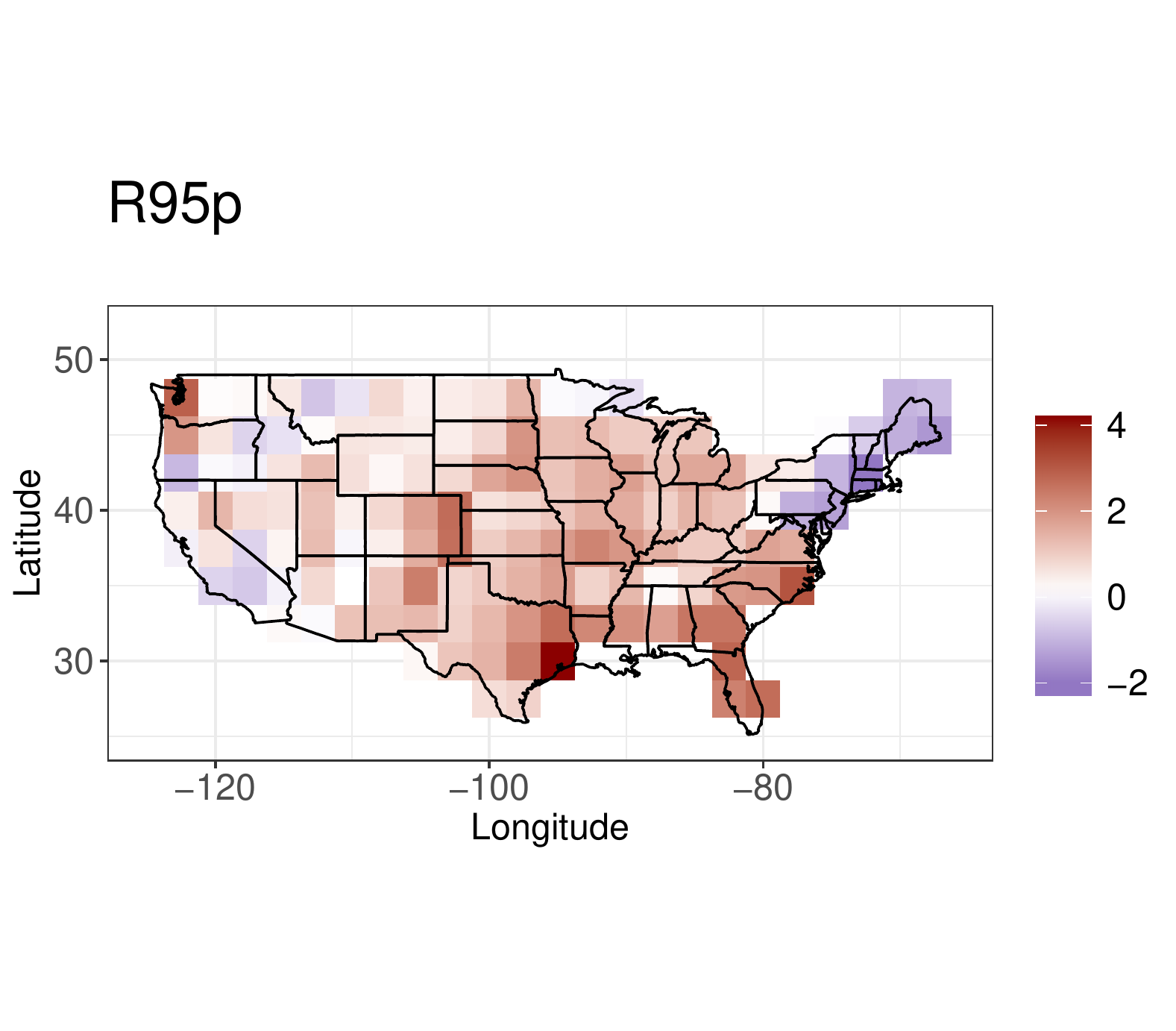} \\
\adjincludegraphics[height = 0.25\linewidth, width=0.42\linewidth, trim = {{.0\width} {.15\width} {.0\width} {.15\width}}, clip]{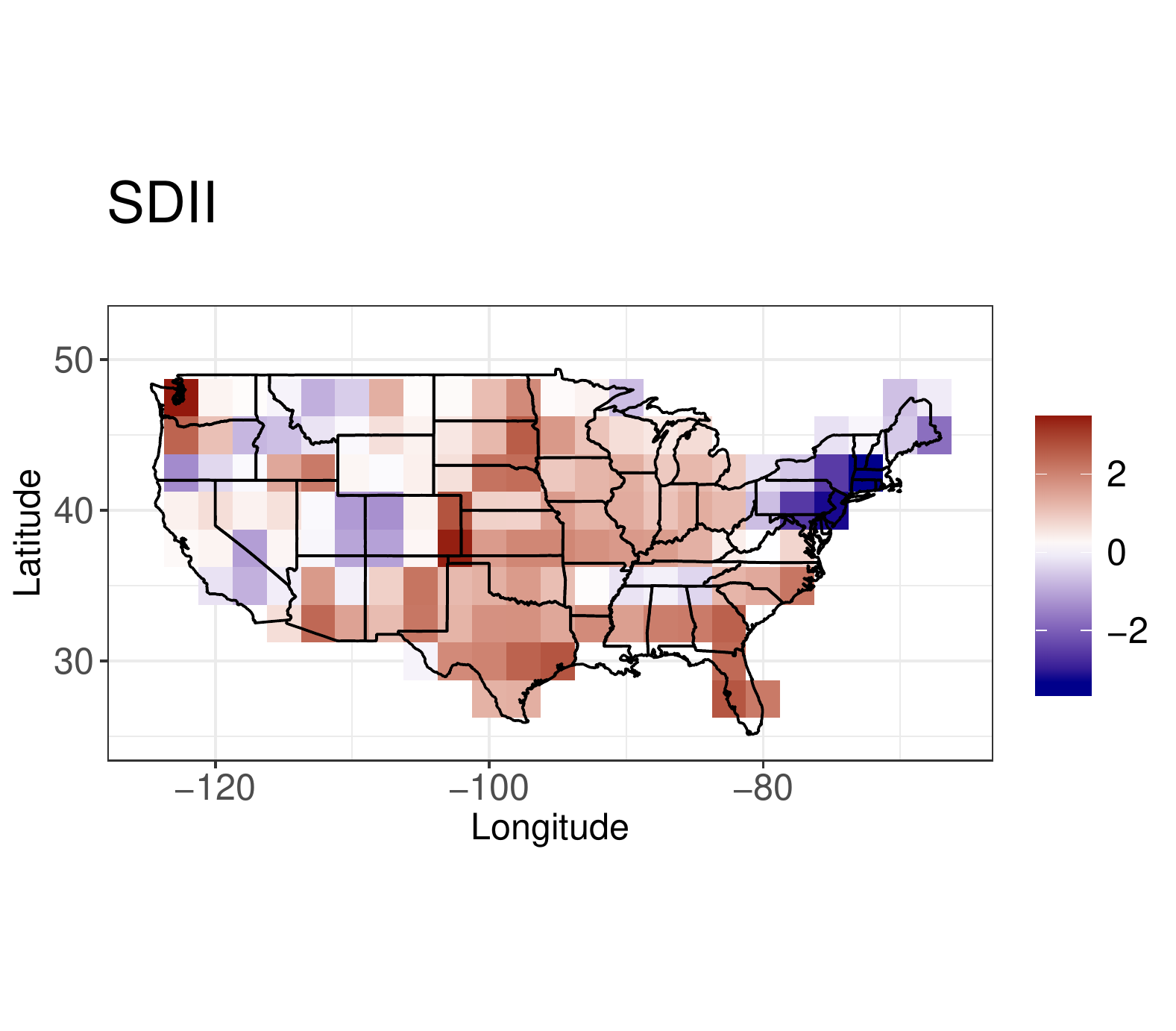}
\adjincludegraphics[height = 0.25\linewidth, width=0.42\linewidth, trim = {{.0\width} {.15\width} {.0\width} {.15\width}}, clip]{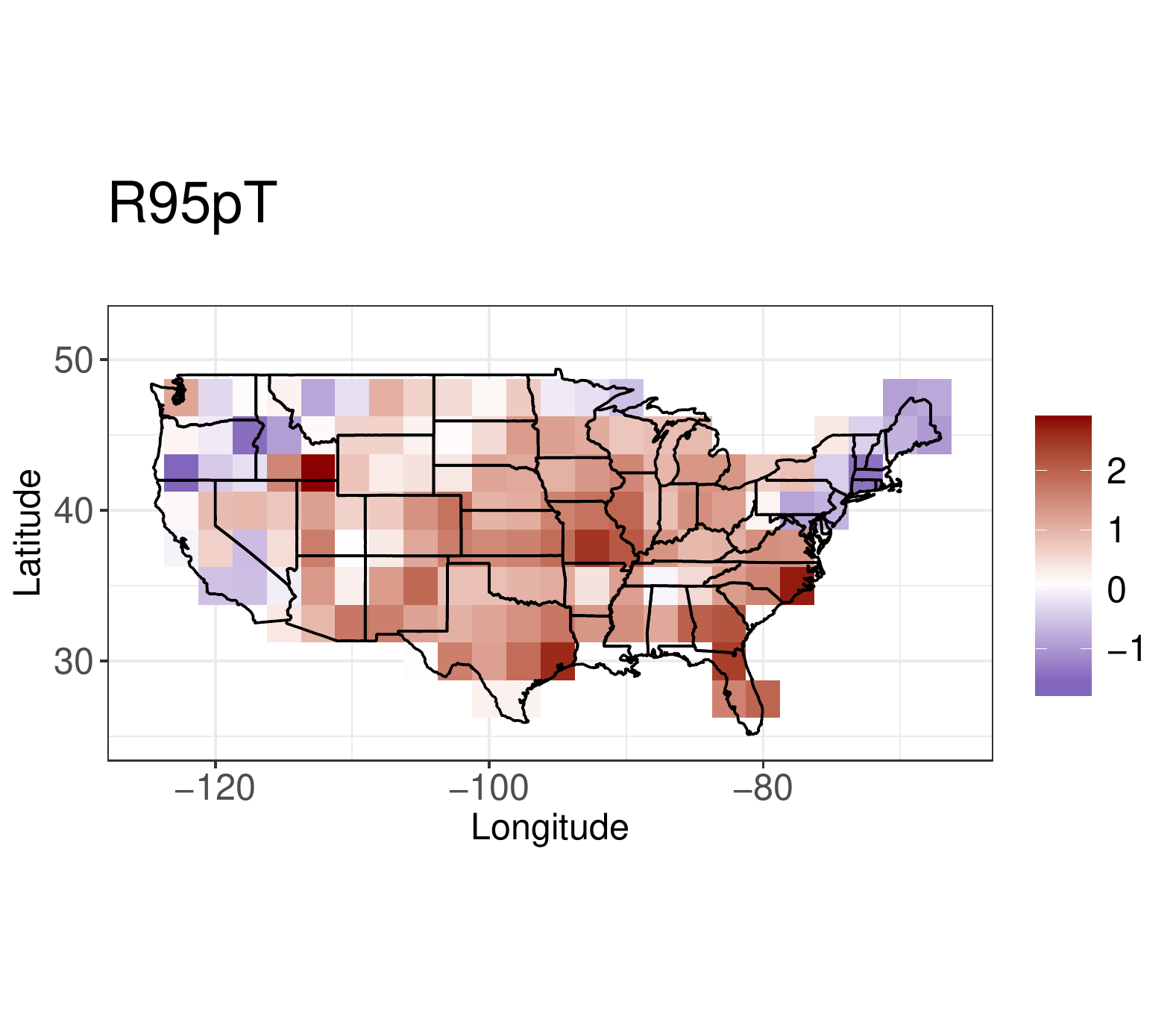} \\
\adjincludegraphics[height = 0.25\linewidth, width=0.42\linewidth, trim = {{.0\width} {.15\width} {.0\width} {.15\width}}, clip]{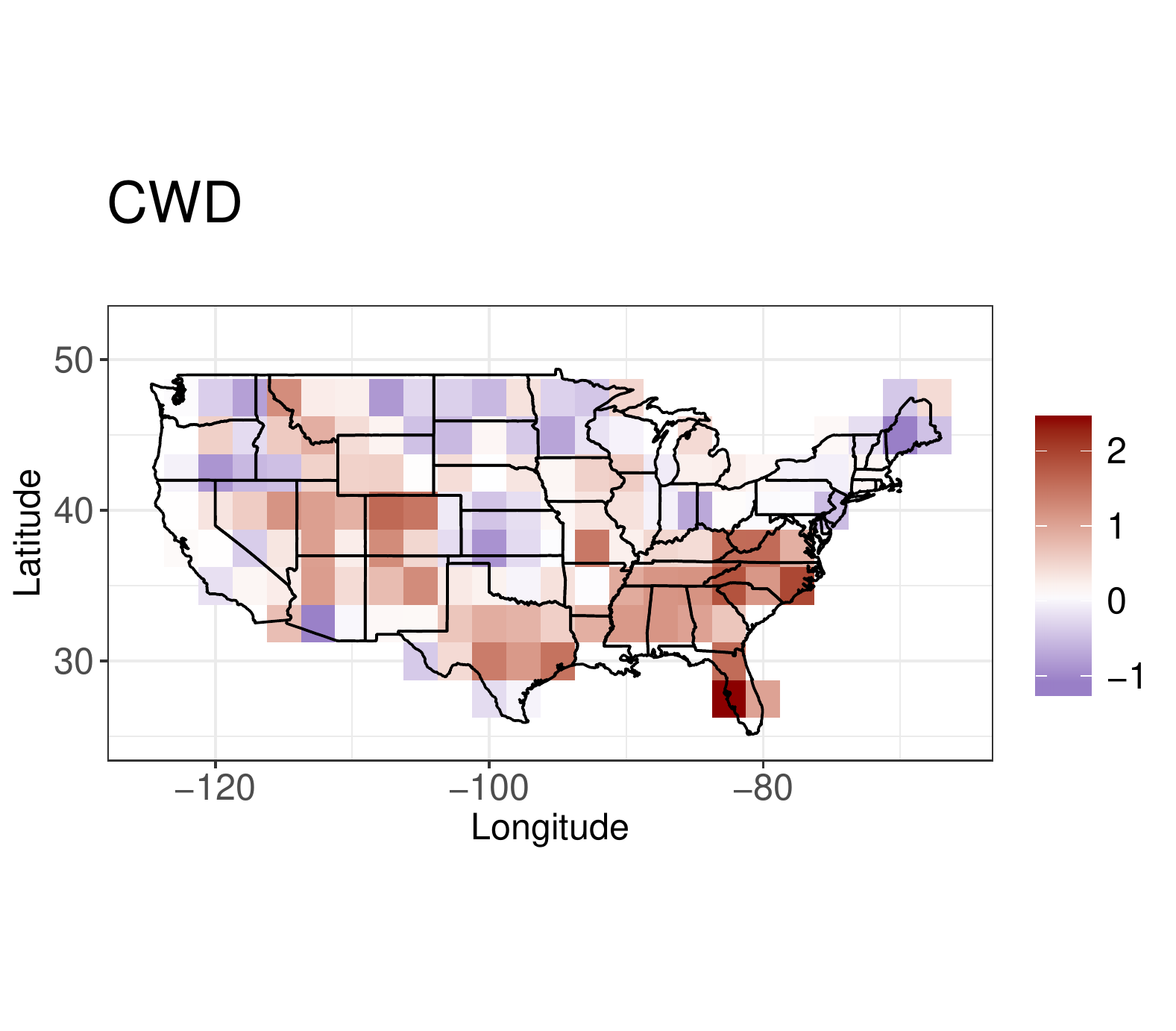}
\adjincludegraphics[height = 0.25\linewidth, width=0.42\linewidth, trim = {{.0\width} {.15\width} {.0\width} {.15\width}}, clip]{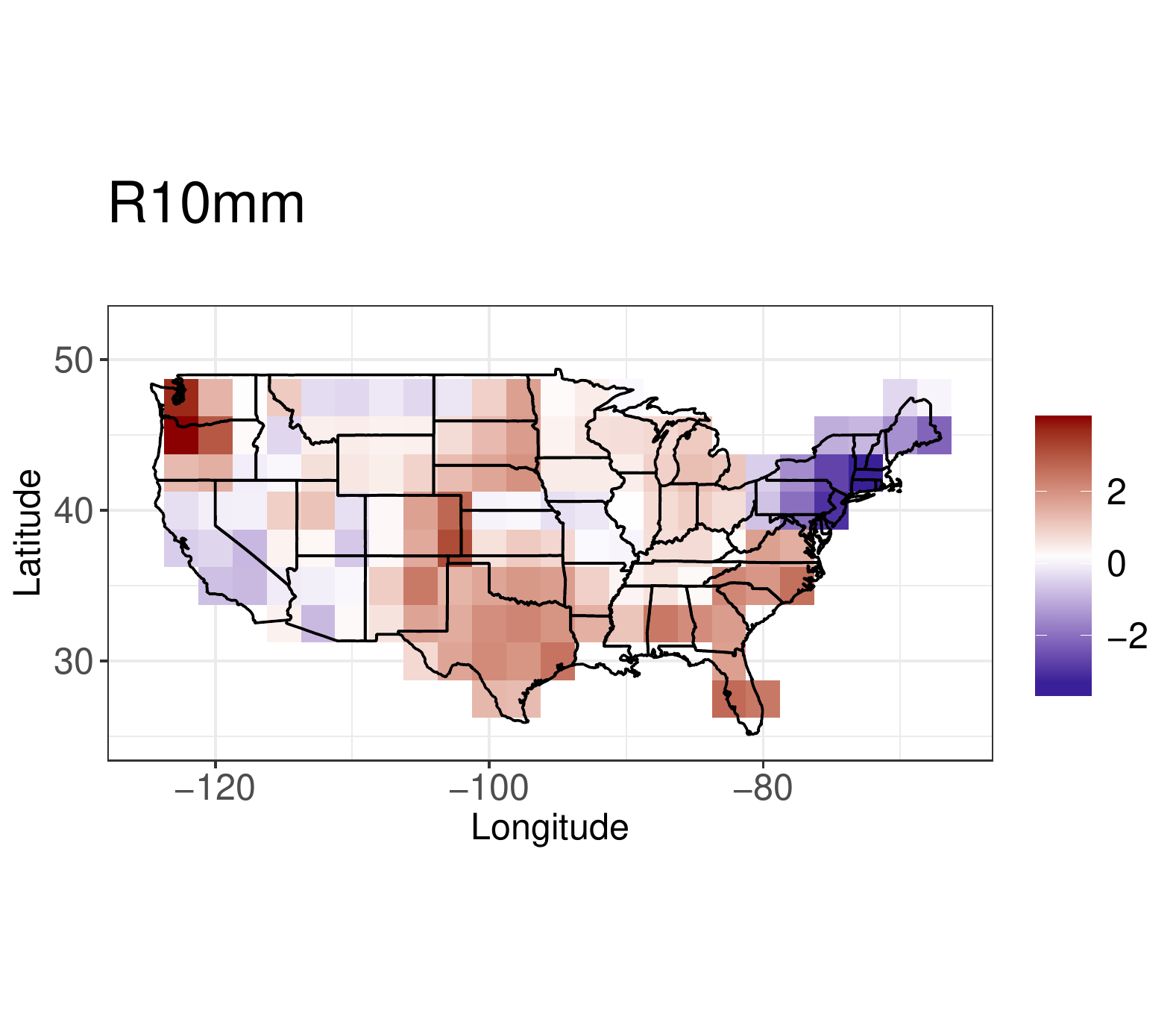} \\
\adjincludegraphics[height = 0.25\linewidth, width=0.42\linewidth, trim = {{.0\width} {.15\width} {.0\width} {.15\width}}, clip]{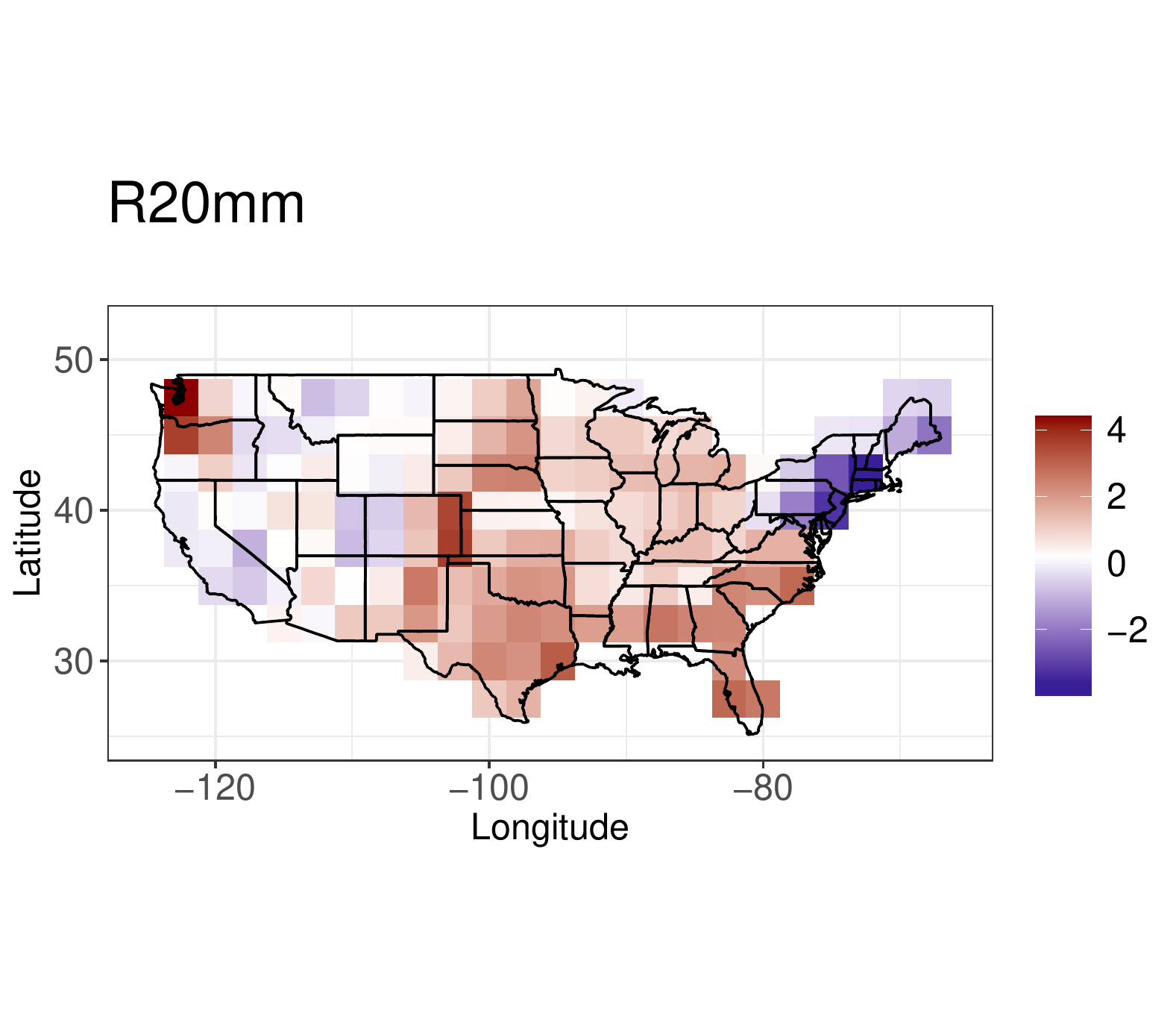}
\adjincludegraphics[height = 0.25\linewidth, width=0.42\linewidth, trim = {{.0\width} {.15\width} {.0\width} {.15\width}}, clip]{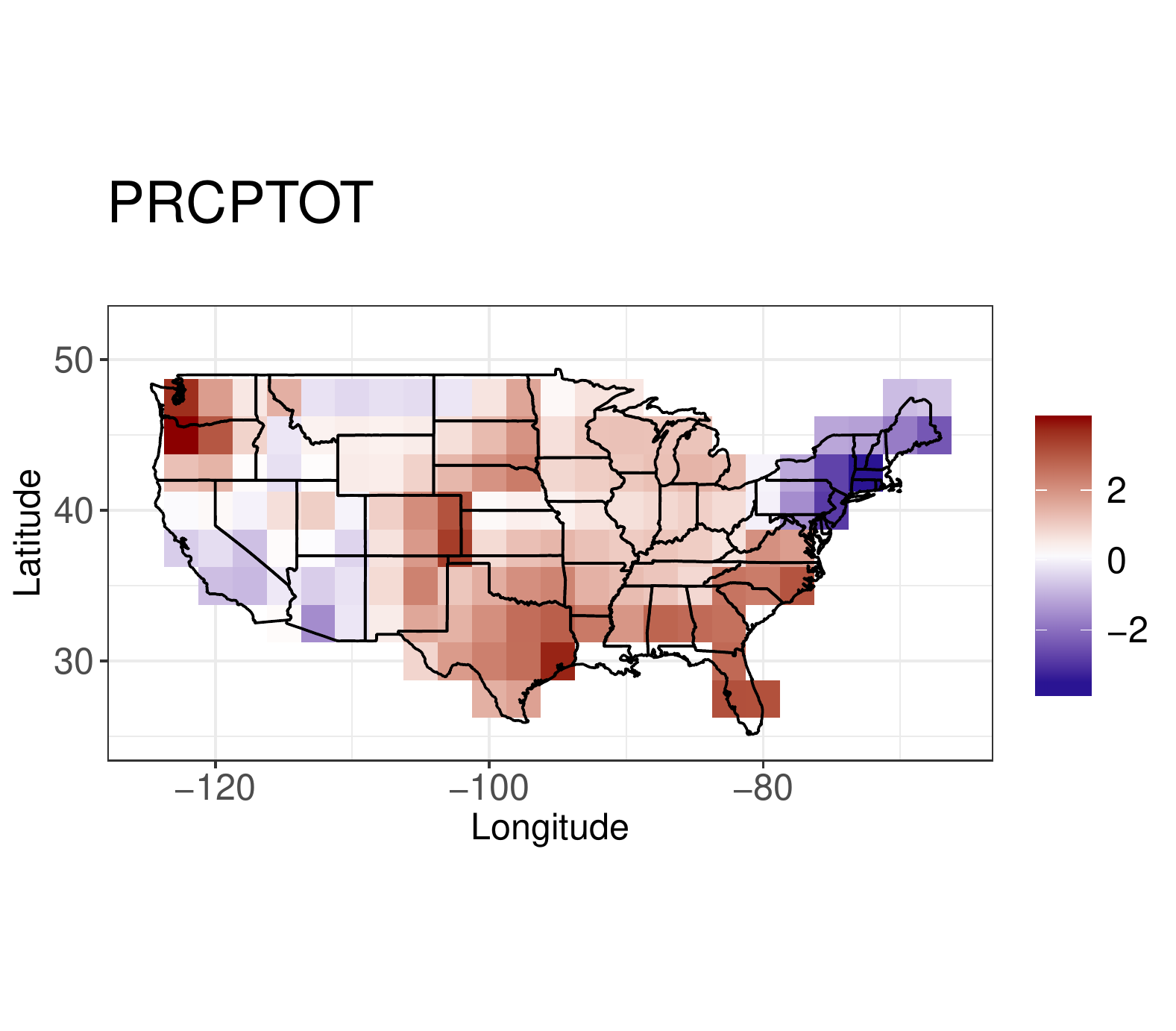}
\caption{Spatial maps of the $t$-statistic values based on fitting multivariate skew-$t$ processes.}
\label{fig_tstats}
\end{figure}

Along the posterior mean of $\Delta_p(\bm{s})$, we calculate the posterior SD and the $t$-values, the ratio of posterior mean and posterior SD. The spatial maps of the $t$-values for MSTP are provided in Figure \ref{fig_tstats}. A value of $t > 2$ is considered to be significant positive change while $t <- 2$ is considered to be significant negative change. There is significant positive change for all the indexes in the southern and south-eastern parts of US. For the regions South-West and West-North-Central, none of the indexes have significant changes except for the single case of significant positive change in mean R95pT in southeast Wyoming. For the indexes SDII, R10mm, R20mmand PRCPTOT, significant negative change is observed near New Jersey and Connecticut. R10mm and R20mm have significant positive increase near Seattle.

\section{Final remarks}
\label{discussions}

In this paper, we propose a multivariate spatial skew-$t$ process model for joint modeling of extreme climate indexes. While univariate spatial extremes have been studied by several authors, models for multivariate spatial extremes are scarce and the existing few models are based on max-stable processes which is computationally demanding. Thus, our method serves the purpose of joint modeling of multivariate spatial extremes with the computational complexity being comparable with multivariate geostatistics approaches. 


There are several possible extensions of the proposed model. Instead of considering the separable covariance structure, non-separable covariance structure can be considered. More details are provided in \cite{gelfand2010handbook}. We create a multivariate skew-$t$ process by mixing random scalar terms in mean and scale. Instead, considering matrix mixing or considering the random vectors for the mean process, a more generalized class of skew-$t$ process models can be constructed.

We analyze several extreme climate indexes and note that there is positive change for all the indexes in the southern and south-eastern parts of US. For the regions South-West and West-North-Central, none of the indexes have significant changes except for the single case of significant positive change in mean R95pT in southeast Wyoming. For the indexes SDII, R10mm, R20mmand PRCPTOT, significant negative change is observed near New Jersey and Connecticut. R10mm and R20mm have significant positive increase near Seattle.

\section*{Appendix}
\label{appendix}

\subsection{MCMC sampling}

Posterior inference about the model parameters have been drawn using Markov chain Monte Carlo procedure implemented in R (\url{http://www.r-project.org}). In case it is possible to consider a conjugate prior, we select it. For some parameters, existences of conjugate prior distributions are unknown. We use random walk Metropolis-Hastings steps to update such parameters. We tune the candidate distributions in Metropolis-Hastings steps during the burn-in period so that the acceptance rate remains in between 0.3 and 0.5.

The set of parameters and hyper-parameters in the model are $\Theta = \left\lbrace \bm{\beta}, \bm{\mu}_{\bm{\beta}}, \bm{\lambda}, \lbrace z_t \rbrace_{t=1}^T, \lbrace \sigma_t^2 \rbrace_{t=1}^T, a, \right.$ $\left. \Sigma_I, \Sigma_B, \rho, \nu, \gamma \right\rbrace$. The MCMC steps for updating the parameters in $\Theta$ are as follows. Corresponding to a parameter (or a set of parameters), by $rest$, we mean the data, all the parameters and hyperparameters in $\Theta$ except that parameter (or that set of parameters).

\noindent \underline{\textbf{$\bm\beta | rest$}} \\
The posterior density of $\bm\beta$ is  $\bm\beta | rest \sim N_p(\bm{\mu}_{\bm{\beta}}^\ast, \Sigma_{\bm{\beta}}^\ast)$ where
\begin{eqnarray}
\nonumber && \Sigma_{\bm{\beta}}^\ast = \Sigma_S \otimes \Sigma_I \otimes \left[ \left(\sum_{t = 1}^{T} \frac{1}{\sigma_t^2} \bm{x}_t \bm{x}'_t  \right) + \Sigma_B^{-1}\right]^{-1} \\
\nonumber && \mu_\beta^\ast =  \Sigma_\beta^\ast \left[ \sum_{t = 1}^{T} \frac{1}{\sigma_t^2} \left[ \left( \Sigma_S^{-1} \otimes \Sigma_I^{-1}  \right) \left(\mathbf{Y}_t - \lvert z_t \rvert \textbf{1}_n \otimes  \bm{\lambda} \right) \right] \otimes \bm{x}_t + \left( \Sigma_S^{-1} \bm{1}_n \right) \otimes \left( \Sigma_I^{-1} \bm{\mu}_{\bm{\beta}} \right) \otimes \left( \Sigma_B^{-1} \bm{1}_L \right) \right].
\end{eqnarray}

\noindent \underline{\textbf{$\bm\mu_{\bm\beta} | rest$}} \\
Suppose $\mathbf{B}$ denotes the $nL \times P$ matrix with the $p$-th column $\mathbf{B}_p = [\bm{\beta}_p(\bm{s}_1)', \ldots, \bm{\beta}_p(\bm{s}_n)']'$. We consider the prior $\bm\mu_{\bm\beta} \sim \textrm{N}(\bm{0}, 100^2 I_P)$. The posterior density of $\bm\mu_{\bm\beta}$ is $\bm\mu_{\bm\beta} | rest \sim N_p(\bar{\bm{\mu}}_{\bm{\beta}}, \Sigma_{\bm\mu_{\bm\beta}})$ where
\begin{eqnarray}
\nonumber && \Sigma_{\bm\mu_{\bm\beta}} = \left[ \left(\bm{1}'_n \Sigma_S^{-1} \bm{1}_n \right) \left(\bm{1}'_L \Sigma_B^{-1} \bm{1}_L \right) \Sigma_I^{-1} + 100^{-2} I_P \right]^{-1} \\
\nonumber && \bar{\bm{\mu}}_{\bm{\beta}} = \Sigma_{\bm\mu_{\bm\beta}} \Sigma_I^{-1} B' \left[ \left( \Sigma_S^{-1} \bm{1}_n \right) \otimes \left( \Sigma_B^{-1} \bm{1}_L \right) \right].
\end{eqnarray}

\noindent \underline{\textbf{$\bm\lambda | rest$}} \\
We consider the prior $\bm\lambda \sim \textrm{N}(\bm{0}, 10^2 I_P)$. The posterior density of $\bm\lambda$ is  $\bm\lambda | rest \sim N_p(\bm{\mu}_{\bm{\lambda}}^\ast, \Sigma_{\bm{\lambda}}^\ast)$ where
\begin{eqnarray}
\nonumber && \Sigma_{\bm\lambda}^\ast = \left[ \left(\bm{1}'_n \Sigma_S^{-1} \bm{1}_n \right) \left( \sum_{t=1}^{T} \frac{z_t^2}{\sigma_t^2} \right) \Sigma_I^{-1} + 10^{-2} I_P \right]^{-1} \\
\nonumber && \bm{\mu}_{\bm{\lambda}}^\ast = \Sigma_{\bm\lambda}^\ast \left[\Sigma_I^{-1} \left( \sum_{t=1}^{T} \frac{\lvert z_t \rvert}{\sigma_t^2} \left[\bm{Y}^*_t - \bm{\mu}^*_t \right]' \right) \left( \Sigma_S^{-1} \bm{1}_n \right) \right].
\end{eqnarray}
where $\bm{Y}^*_t$ and $\bm{\mu}^*_t$ are $n \times P$ matrices with the $p$-th columns are $[Y_{tp}(\bm{s}_1), \ldots, Y_{tp}(\bm{s}_n)]'$ and $[\mu_{tp}(\bm{s}_1), \ldots, \mu_{tp}(\bm{s}_n)]'$ respectively. 

\noindent \underline{\textbf{$\lvert z_t \rvert | rest$}} \\
	As $z_t$ is not identifiable, we treat $\lvert z_t \rvert$ as a parameter and update within the MCMC steps. Here $\lvert z_t \rvert \sim HN(\sigma_t^2)$ where $HN$ denotes the half-normal density. The posterior density of $\lvert z_t \rvert$ conditioned on $rest$ is given by
	\begin{eqnarray}
	\nonumber f(\lvert z_t \rvert | rest) \propto \exp\left[-\frac{1}{2} \frac{1}{\sigma_t^2} \left(\bm{Y}_t - \bm{\mu}_t - \lvert z_t \rvert \textbf{1}_n \otimes  \bm{\lambda} \right)' \Sigma_S^{-1} \otimes \Sigma_I^{-1} \left(\bm{Y}_t - \bm{\mu}_t - \lvert z_t \rvert \textbf{1}_n \otimes  \bm{\lambda} \right) -\frac{1}{2} \frac{\lvert z_t \rvert^2}{\sigma_t^2}  \right] I(\lvert z_t \rvert > 0)
	\end{eqnarray}
	i.e., $\lvert z_t \rvert | rest \sim N_{(0, \infty)}\left(\mu_z^*, \sigma_z^{*2}\right)$ where
\begin{eqnarray}
\nonumber && \sigma_z^{*2} = \sigma_t^2 \left[ 1 + \left(\bm{1}'_n \Sigma_S^{-1} \bm{1}_n \right) \left(\bm{\lambda}' \Sigma_I^{-1} \bm{\lambda} \right) \right]^{-1}, \\
\nonumber && \mu_z^* = \left[ 1 + \left(\bm{1}'_n \Sigma_S^{-1} \bm{1}_n \right) \left(\bm{\lambda}' \Sigma_I^{-1} \bm{\lambda} \right) \right]^{-1} \left[ \bm{1}'_n \Sigma_S^{-1} \left[\bm{Y}^*_t - \bm{\mu}^*_t \right] \Sigma_{I}^{-1} \bm{\lambda} \right],
\end{eqnarray}
where $\bm{Y}^*_t$ and $\bm{\mu}^*_t$ are $n \times P$ matrices as defined for calculating the posterior density of $\bm\lambda$.

\noindent \underline{\textbf{$\sigma_t^2 | rest$}} \\
The posterior density of $\sigma_t^2$ given $rest$ is
	\begin{eqnarray}
	\nonumber && \sigma_t^2 | rest \sim IG \left(\frac{a + n P + 1}{2}, \frac{a + (\bm{Y}_t - \bm{\mu}_t - \lvert z_t \rvert \textbf{1}_n \otimes  \bm{\lambda})' \Sigma_S^{-1} \otimes \Sigma_I^{-1} (\bm{Y}_t - \bm{\mu}_t - \lvert z_t \rvert \textbf{1}_n \otimes  \bm{\lambda}) + z_t^2}{2} \right).
	\end{eqnarray}

\noindent \underline{\textbf{$a | rest$}} \\
	We consider discrete uniform prior for $a$, i.e., $a \overset{iid}{\sim} DU(0.1, 0.2, \ldots, 19.9, 20.0)$. The posterior distribution of $a$ given $rest$ is 
	\begin{eqnarray}
	\nonumber Pr(a = a^\ast | rest) \propto \prod_{t = 1}^{T} f_{IG}(\sigma_t^2; a^\ast / 2, a^\ast / 2)
	\end{eqnarray}
where $f_{IG}$ denotes the inverse gamma density. We draw random sample from the discrete support $\lbrace 0.1, 0.2, \ldots, 19.9, 20.0 \rbrace$ with probabilities proportional to $Pr(a = a^\ast | rest)$.

\noindent \underline{{$\Sigma_I | rest$}} \\
We consider the prior $\Sigma_I \sim \textrm{IW}(0.01, 0.01 I_P)$. The posterior density of $\Sigma_I$ given $rest$ is $\textrm{IW}(\nu_{I}^*, \Psi_{I}^*)$ where
\begin{eqnarray}
\nonumber \nu_{I}^* &=& 0.01 + n T + n L \\
\nonumber \Psi_{I}^* &=& 0.01 I_P + \sum_{t = 1}^{T} (\bm{Y}_t^* - \bm{\mu}_t^* - \lvert z_t \rvert \textbf{1}_n \otimes  \bm{\lambda}')' \Sigma_S^{-1} (\bm{Y}_t^* - \bm{\mu}_t^* - \lvert z_t \rvert \textbf{1}_n \otimes  \bm{\lambda}') \\
\nonumber && + \left(\mathbf{B} - \bm{1}_{n L} \otimes \bm\mu_{\bm\beta}' \right)' \Sigma_S^{-1} \otimes \Sigma_B^{-1} \left(\mathbf{B} - \bm{1}_{n L} \otimes \bm\mu_{\bm\beta}' \right)
	\end{eqnarray}
where $\bm{Y}^*_t$ and $\bm{\mu}^*_t$ are $n \times P$ matrices as defined for calculating the posterior density of $\bm\lambda$ and $\mathbf{B}$ is as defined for calculating the posterior density of $\bm\mu_{\bm\beta}$.

\noindent \underline{{$\Sigma_B | rest$}} \\
We consider the prior $\Sigma_B \sim \textrm{IW}(0.01, 0.01 I_L)$. The posterior density of $\Sigma_B$ given $rest$ is
\begin{eqnarray}
\nonumber \Sigma_B| rest &\sim& \textrm{IW}\left(0.01 + n P, 0.01 I_L + \left(\mathbf{B}^* - \bm{1}_{L}' \otimes  \bm{1}_{n} \otimes \bm\mu_{\bm\beta} \right)' \Sigma_S^{-1} \otimes \Sigma_I^{-1} \left(\mathbf{B}^* - \bm{1}_{L}' \otimes  \bm{1}_{n} \otimes \bm\mu_{\bm\beta} \right) \right)
\end{eqnarray}
where $\mathbf{B}^*$ denotes the $nP \times L$ matrix with the $l$-th column $\mathbf{B}^*_l = [\bm{\beta}_l^*(\bm{s}_1)', \ldots, \bm{\beta}_l^*(\bm{s}_n)']'$ and $\bm{\beta}_l^*(\bm{s}_i) = [\beta_{l1}(\bm{s}_i), \ldots, \beta_{lP}(\bm{s}_i)]'$.

\noindent \underline{\textbf{$\rho, \nu, \gamma | rest$}} \\
The parameters are updated using Metropolis-Hastings algorithm. Here we update the two parameters $\rho$ and $\nu$ together (due to strong negative correlation of the joint posterior density) and separately we update $\gamma$. We consider the priors $\rho \sim \textrm{U}(0, D)$, $\nu^* = \log[\nu] \sim N(-1.2, 1^2)$ and $\gamma \sim \textrm{U}(0, 1)$ where $D$ denotes the maximum distance between two grid points (in degrees) within US. We update $\nu$ in the log scale.  

Suppose $\rho^{(m)}$ denotes the MCMC sample from $\rho$ at the $m$-th MCMC iteration. Considering a logit transformation, we obtain $\rho^{(m)*} \in \Re$ and generate a sample $\rho^{(c)*} \sim N\left( \rho^{(m)*}, s_{\rho}^2 \right)$. Subsequently, using an inverse-logit transformation, we obtain $\rho^{(c)}$ from $\rho^{(c)*}$. Further, we generate $\nu^{*(c)} \sim N\left( \nu^{*(m)}, s_{\nu}^2 \right)$ where $\nu^{*(m)}$ denotes the MCMC sample from $\nu^*$ at the $m$-th MCMC iteration. Suppose $\Sigma_S^{(m)}$ denotes the spatial correlation matrix based on $\left(\rho^{(m)}, \nu^{(m)}, \gamma^{(m)}\right)$ and $\Sigma^{(c)}$ denotes the spatial correlation matrix based on $\left(\rho^{(c)}, \nu^{(c)}, \gamma^{(m)}\right)$.
The acceptance ratio is 
\begin{eqnarray}
\nonumber R &=& \frac{\prod_{t = 1}^{T} N_{nP}\left( \mathbf{Y}_{t}; \bm\mu_t + \lvert z_t \rvert \textbf{1}_n \otimes  \bm{\lambda} , \sigma_t^2 \Sigma_{S}^{(m)}  \otimes \Sigma_I \right) }{\prod_{t = 1}^{T} N_{nP}\left( \mathbf{Y}_{t}; \bm\mu_t + \lvert z_t \rvert \textbf{1}_n \otimes  \bm{\lambda}, \sigma_t^2 \Sigma_{S}^{(c)} \otimes \Sigma_I \right) } \times \frac{f(\nu^{*(c)})}{f(\nu^{*(m)})} \times \frac{\rho^{(c)} \left(D - \rho^{(c)}\right)}{\rho^{(m)} \left(D - \rho^{(m)}\right)}.
\end{eqnarray}
where $f(\nu^{*})$ denotes the prior density of $\nu^*$. The candidates are accepted with probability $min \lbrace R,1 \rbrace$. While updating the parameter $\gamma$, the candidates are generated similar to $\rho$.

\subsection{Marginal distributions}

The marginal densities of univariate and multivariate skew-$t$ distributions are provided in the following.\\
\noindent \underline{Univariate skew-$t$ distribution}
The density function of $Y_{tp}(\bm{s}_i)$ is
	$$f_{Y_{tp}(\bm{s}_i)}(y) = \frac{2}{\sqrt{\Sigma_I^{(p,p)} + \lambda_p^2}} f_T\left(\frac{y - \mu_{tp}(\bm{s}_i)}{\sqrt{\Sigma_I^{(p,p)} + \lambda_p^2}}; a \right) F_T\left( \frac{\lambda_p}{\sqrt{\Sigma_I^{(p,p)}}} \frac{y - \mu_{tp}(\bm{s}_i)}{\sqrt{\Sigma_I^{(p,p)} + \lambda_p^2}} \sqrt{\frac{a + 1}{a + \frac{(y - \mu_{tp}(\bm{s}_i))^2}{\Sigma_I^{(p,p)} + \lambda_p^2}}}; a + 1 \right) $$
	where $f_T(\cdot; a)$ and $F_T(\cdot;a)$ are density and cumulative distribution functions (CDF) of univariate Student's $t$ distribution (location = 0 and scale = 1) with $a$ degrees of freedom and $\Sigma_I^{(p,p)}$ denotes the $(p,p)$-th element of $\Sigma_I$.
	
\noindent \underline{Multivariate skew-$t$ distribution} \\
The density function of $\mathbf{Y}_t(\bm{s}_i)$ is
	$$f_{\mathbf{Y}_t(\bm{s}_i)}(\bm{y}) = 2 f_{T_P}\left(\bm{z}; \bm{0}, \Sigma_Y, a \right) F_T\left( \frac{\bm{\lambda}' \Sigma_I^{-1} \bm{z}}{\sqrt{1 + \bm{\lambda}'\Sigma_I^{-1} \bm{\lambda}}} \sqrt{\frac{a + P}{a + \bm{z}'\Sigma_Y^{-1} \bm{z}}}; a + P \right) $$
	where $\bm{z} = \bm{y} - \bm{\mu}_t(\bm{s})$, $f_{T_P}(\cdot; \bm{\mu}, \Sigma_Y, a)$ is the density function of $P$-variate Student's $t$ distribution with location $\bm{\mu}$, shape matrix $\Sigma_Y$ and $a$ degrees of freedom and $F_T(\cdot;a)$ is the CDF of univariate Student's $t$ distribution (location = 0 and scale = 1) with $a$ degrees of freedom. The matrix $\Sigma_Y$ is given by $\Sigma_Y = \Sigma_I + \bm{\lambda} \bm{\lambda}' $.

\bibliographystyle{imsart-nameyear}
\bibliography{spatialextremes}

\end{document}